\documentclass[extra]{gji}  %>>> use for US letter paper
\usepackage[]{graphicx,bbold}
\usepackage[]{url}
\usepackage{amssymb,amsmath}
\newcommand{\dXlm}{X'_{lm}}
\newcommand{\dXlpm}{X'_{l'm}}
\newcommand{\renormprod}{\sqrt{l(l+1)\hsom l'(l'+1)}} %
\newcommand{\sumalpha}{\sum_{\alpha=1}^{3(L+1)^2-2}}
\newcommand{\intoth}{\int_0^\Theta}
\newcommand{\sindth}{\sin \theta \, d\theta}

\newcommand{\flmP}{f_{lm}^P}
\newcommand{\flmB}{f_{lm}^B}
\newcommand{\flmC}{f_{lm}^C}
\newcommand{\glmP}{g_{lm}^P}
\newcommand{\glmB}{g_{lm}^B}
\newcommand{\glmpC}{g_{l'm'}^C}
\newcommand{\glmpP}{g_{l'm'}^P}
\newcommand{\glmpB}{g_{l'm'}^B}
\newcommand{\glmC}{g_{lm}^C}

\newcommand{\renormsing}{\sqrt{l(l+1)}}
\newcommand{\thvec}{\boldsymbol{\hat\theta}}
\newcommand{\phvec}{\boldsymbol{\hat\phi}}
\newcommand{\bPlm}{{\bf P}_{lm}}
\newcommand{\bBlm}{{\bf B}_{lm}}
\newcommand{\bClm}{{\bf C}_{lm}}
\newcommand{\bPlmp}{{\bf P}_{l'm'}}
\newcommand{\bBlmp}{{\bf B}_{l'm'}}
\newcommand{\bClmp}{{\bf C}_{l'm'}}
\newcommand{\divsin}{(\sin \theta)^{-1}}
\newcommand{\divsinsq}{(\sin \theta)^{-2}}
 
\newcommand{\Poo}{{\bf P}_{00}}

\newcommand{\hsom}{\hspace{0.1em}}

\newcommand{\WS}{^{\mathrm{WS}}}
\newcommand{\SP}{^{\mathrm{SP}}}
\newcommand{\MT}{^{\mathrm{MT}}}

\newcommand{\pinv}{{\mathrm{pinv}}}
\newcommand{\bxi}{\mbox{\boldmath$\xi$}}
\newcommand{\bdelta}{\mbox{\boldmath$\delta$}}
\newcommand{\intinf}{\int_{-\infty}^{\infty}}

\newcommand{\intW}{\int_{-W}^{W}}
\newcommand{\sumshLp}{\suml_{l'=0}^{L}\suml_{m'=-l'}^{l'}}
\newcommand{\D}{^{\mathrm{1D}}}
\newcommand{\DD}{^{\mathrm{2D}}}
\newcommand{\DDD}{^{\mathrm{3D}}}
\newcommand{\DDDD}{^{\mathrm{vec}}}
\newcommand{\Dxxp}{D(\bx,\bx')}
\newcommand{\Dkkp}{D(\bk,\bk')}
\newcommand{\Dttp}{D(t,t')}

\newcommand{\Dxixip}{D(\bxi,\bxi')}
\newcommand{\Dxixi}{D(\bxi,\bxi)}
\newcommand{\fnorm}{(2\pi)^{-2}}
\newcommand{\jnorm}{(2\pi)^{-1}}

\newcommand{\intK}{\int_\mathcal{K}}

\newcommand{\bx}{\mathbf{x}}
\newcommand{\bga}{{\bf g}_{\alpha}}
\newcommand{\bgb}{{\bf g}_{\beta}}
\newcommand{\bef}{\mathbf{f}}

\newcommand{\bk}{\mathbf{k}}
\newcommand{\dbk}{\,d\bk}

\newcommand{\dth}{\,d \theta}
\newcommand{\plphi}{\partial_\phi}
\newcommand{\plth}{\partial_\theta}
\newcommand{\dbx}{\,d\bx}

\newcommand{\domg}{\,d\Omega} 
\newcommand{\fracd}[2]{\frac{\displaystyle{#1}}{\displaystyle{#2}}} 
\newcommand{\intR}{\int_\mathcal{R}}
\newcommand{\intr}{\int_R} 
\newcommand{\intbr}{\int_{\bar{R}}} 
 
\newcommand{\into}{\int_\Omega}

\newcommand{\also}{\quad\mbox{and}\quad} 
\newcommand{\with}{\quad\mbox{with}\quad} 
 
\newcommand{\for}{\quad\mbox{for}\quad} 
\newcommand{\dab}{\delta_{\alpha\beta}}

\newcommand{\dllp}{\delta_{ll'}} 
\newcommand{\dmmp}{\delta_{mm'}} 
\newcommand{\Ylm}{Y_{lm}} 
\newcommand{\Ylmrh}{Y_{lm}(\rhat)} 
\newcommand{\Ylmrhp}{Y_{lm}(\rhat')}

\newcommand{\Ylmp}{Y_{l'm'}} 
\newcommand{\Xlm}{X_{lm}} 
\newcommand{\Xlamth}{X_{l |m|}(\theta)}

\newcommand{\Xlmth}{X_{lm}(\theta)} 
\newcommand{\Xlmthp}{X_{lm}(\theta')} 
\newcommand{\Xlpm}{X_{l'm}}

\newcommand{\Dlmlm}{D_{lm,lm}}
\newcommand{\Dlmlmp}{D_{lm,l'm'}}
\newcommand{\Blmlmp}{B_{lm,l'm'}}
\newcommand{\Blmlm}{B_{lm,lm}}
\newcommand{\Clmlmp}{C_{lm,l'm'}}
\newcommand{\Clmlm}{C_{lm,lm}}
\newcommand{\bDlmlmp}{\bar{D}_{lm,l'm'}}

\newcommand{\diag}{\textrm{diag}} 
 
\newcommand{\Drhrhp}{D(\rhat,\rhat')} 
 
\newcommand{\bDrhrhp}{\Db(\rhat,\rhat')} 
 
\newcommand{\grhp}{g(\rhat')} 
\newcommand{\garh}{g_\alpha(\rhat)}

\newcommand{\glma}{g_{lm \hsp\alpha}} 
 
\newcommand{\galm}{\glma} 
 
\newcommand{\glmb}{g_{lm\hsp\beta}} 
\newcommand{\glmpb}{g_{l'm'\hsp\beta}} 
\newcommand{\gblmp}{\glmpb} 
 
\newcommand{\Lpot}{(L+1)^2} 
\newcommand{\sumapot}{\sum_{\alpha=1}^{\Lpot}} 
 
\newcommand{\sumakR}{\sum_{\alpha>N\DDD}^{\Lpot}} 
\newcommand{\sumaN}{\sum_{\alpha=1}^{N\DDD}}

\newcommand{\suml}{\sum\limits}

\newcommand{\sumsh}{\suml_{l=0}^{\infty}\suml_{m=-l}^{l}} 
\newcommand{\sumshL}{\suml_{l=0}^{L}\suml_{m=-l}^{l}}
\newcommand{\summ}{\suml_{m=-l}^{l}}
\newcommand{\tlofp}{\left(\frac{2l+1}{4\pi}\right)} 
\newcommand{\Plm}{P_{lm}}

\newcommand{\bnabla}{\mbox{\boldmath$\nabla$}} 
\newcommand{\blambda}{\mbox{\boldmath$\Lambda$}} 
\newcommand{\bsigma}{\mbox{\boldmath$\Sigma$}} 
 
\newcommand{\rhat}{\mbf{\hat{r}}} 
\newcommand{\shat}{\hat{s}} 
\newcommand{\Shat}{\hat{S}} 
 
\newcommand{\bmp}{\begin{minipage}}
\newcommand{\emp}{\end{minipage}}
\newcommand{\be}{\begin{equation}} 
\newcommand{\ee}{\end{equation}} 
\newcommand{\ber}{\begin{eqnarray}} 
\newcommand{\eer}{\end{eqnarray}} 
\newcommand{\barray}{\begin{array}} 
\newcommand{\earray}{\end{array}} 
\newcommand{\mbf}{\mathbf}

\newcommand{\nnr}{\nonumber}
\newcommand{\sst}{\scriptstyle} 

\newcommand{\rar}{\rightarrow} 
\newcommand{\ssec}{\subsection} 
\newcommand{\sssec}{\subsubsection} 
 
\newcommand{\hsp}{\hspace*{0.1em}}
\newcommand{\hsps}{\hspace*{0.75em}}
\newcommand{\hspm}{\hspace*{-0.1em}}

\newcommand{\beff}{\mbox{\boldmath$\mathsf{f}$}}
\newcommand{\beg}{\mbox{\boldmath$\mathsf{g}$}}
\newcommand{\bega}{\mbox{\boldmath$\mathsf{g}$}_\alpha}
\newcommand{\begb}{\mbox{\boldmath$\mathsf{g}$}_\beta}
\newcommand{\bg}{\mathbf{g}}
\newcommand{\begg}{\mbox{\boldmath$\mathsf{G}$}}
\newcommand{\beggbar}{\mbox{\boldmath$\bar{\mathsf{G}}$}}
\newcommand{\beggubar}{\mbox{\boldmath{\b{$\mathsf{G}$}}}}
\newcommand{\blambdaubar}{\mbox{\boldmath{\b{$\Lambda$}}}} 
\newcommand{\blambdabar}{\mbox{\boldmath$\bar{\Lambda}$}} 
\newcommand{\Cbubar}{\mbox{\boldmath{\b{$\mathrm{B}$}}}}
\newcommand{\Wbubar}{\mbox{\boldmath{\b{$\mathrm{V}$}}}}
\newcommand{\Ibubar}{\mbox{\boldmath{\b{$\mathrm{I}$}}}}
\newcommand{\bY}{\mbox{\boldmath$\mathsf{Y}$}}
\newcommand{\sK}{\mbox{\boldmath$\mathsf{K}$}}
\newcommand{\sP}{\mbox{\boldmath$\mathsf{D}$}}
\newcommand{\sC}{\mbox{\boldmath$\mathsf{C}$}}
\newcommand{\so}{\mbox{\boldmath$\mathsf{0}$}}
\newcommand{\sB}{\mbox{\boldmath$\mathsf{B}$}}
\newcommand{\bG}{\mbox{\boldmath$\mathsf{G}$}}
\newcommand{\br}{\rhat}
\newcommand{\Tit}{^{\it{\sst{T}}}}
\newcommand{\Trm}{^{\mathrm{\sst{T}}}}
\newcommand{\Tsf}{^\mathsf{T}}
\newcommand{\T}{^{\sf{\sst{T}}}}
\newcommand{\fb}{\mathbf{f}}
\newcommand{\fbh}{\mathbf{\hat{f}}}
\newcommand{\tb}{\mathbf{t}}
\newcommand{\tbh}{\mathbf{\hat{t}}}
\newcommand{\Gb}{\mathbf{G}}
\newcommand{\Ub}{\mathbf{U}}
\newcommand{\Db}{\mathbf{D}}
\newcommand{\Ib}{\mathbf{I}}
\newcommand{\Ibbar}{\bar{\mathbf{I}}}
\newcommand{\Hb}{\mathbf{H}}
\newcommand{\Vb}{\mathbf{V}}
\newcommand{\Bb}{\mathbf{B}}

\newcommand{\Cbbar}{\mathbf{\bar{B}}}

\newcommand{\Wbbar}{\mathbf{\bar{V}}}
\newcommand{\sumshortL}{\suml_{lm}^{L}}
\newcommand{\sumshortLp}{\suml_{l'm'}^{L}}
\newcommand{\bd}{\mbox{\boldmath$\mathsf{d}$}}

\newcommand{\tr}{{\mathrm{tr}}}

\newcommand{\bP}{\mbox{\boldmath$\mathsf{P}$}}

\begin{document} 
\onecolumn
\title[Scalar and Vector Slepian Functions]{Scalar and Vector Slepian
  Functions, Spherical Signal Estimation and Spectral Analysis}
\author[Simons and Plattner]{Frederik J.~Simons$^*$ and Alain Plattner\\
Department of Geosciences, Princeton University, Guyot Hall,
Princeton, NJ, USA\\
$^*$Also with the Program in Applied and Computational Mathematics,
Princeton University, Princeton, NJ, USA
}
\maketitle 
%%%%%%%%%%%%%%%%%%%%%%%%%%%%%%%%%%%%%%%%%%%%%%%%%%%%%%%%%%%%% 
\begin{summary}
It is a well-known fact that mathematical functions
that are timelimited (or spacelimited) cannot be simultaneously
bandlimited (in frequency). Yet the finite precision of
measurement and computation unavoidably bandlimits our observation and
modeling scientific data, and we often only have access to, or are
only interested in, a study area that is temporally or spatially
bounded. In the geosciences we may be interested in spectrally
modeling a time series defined only on a certain interval, or we may
want to characterize a specific geographical area observed using an
effectively bandlimited measurement device. It is clear that analyzing
and representing scientific data of this kind will be facilitated if a
basis of functions can be found that are ``spatiospectrally''
concentrated, i.e. ``localized'' in both domains at the same
time. Here, we give a  theoretical overview of one particular
approach to this ``concentration'' problem, as originally proposed for
time series by Slepian and coworkers, in the 1960s. We show how this
framework leads to practical algorithms and statistically performant
methods for the analysis of signals and their power spectra in one and
two dimensions, and, particularly for applications in the geosciences,
for scalar and vectorial signals defined on the surface of a unit sphere. 
\end{summary}


\section{Introduction}

It is well appreciated that functions cannot have
finite support in the temporal (or spatial) and spectral domain at the
same time~\cite[]{Slepian83}. Finding and representing signals  that
are optimally concentrated in both is a fundamental problem in
information theory  which was solved in the early 1960s by Slepian,  
Landau and Pollak~\cite[]{Slepian+61,Landau+61,Landau+62}. The
extensions and generalizations of this 
problem~\cite[]{Daubechies88a,Daubechies+88,Cohen89,Daubechies90} have
strong connections with the  burgeoning field of wavelet
analysis.  In this contribution, however, we shall \textit{not} talk
about wavelets, the scaled translates of a ``mother'' with vanishing
moments, the tool for multi-resolution
analysis~\cite[]{Daubechies92,Flandrin98,Mallat98}. Rather, we devote our
attention entirely to what we shall collectively refer to as ``Slepian
functions'', in multiple Cartesian dimensions and on the sphere.   

These we understand to be orthogonal families of functions that are
all defined on a common, e.g. geographical, domain, where they are
either optimally concentrated or within which they are exactly
limited, and which at the same time are exactly confined within a
certain bandwidth, or maximally concentrated therein. The measure of
concentration is invariably a quadratic energy ratio, which, though
only one choice out of
many~\cite[]{Donoho+89,Freeden+97,Riedel+95,Freeden+2010,Michel2010} is
perfectly 
suited to the nature of the problems we are attempting to address.
These are, for example: How do we make estimates of signals that are
noisily and incompletely observed? How do we analyze the properties of
such signals efficiently, and how can we represent them economically?
How do we estimate the power spectrum of noisy and incomplete data?
What are the particular constraints imposed by dealing with
potential-field signals (gravity, magnetism, etc) and how is the
altitude of the observation point, e.g. from a satellite in orbit,
taken into account? What are the statistical properties of the
resulting signal and power spectral estimates?

These and other questions have been studied extensively in one
dimension, that is, for time series, but until the twenty-first
century, remarkably little work had been done in the Cartesian plane
or on the surface of the sphere. For the geosciences, the latter two
domains of application are nevertheless vital for the obvious reasons
that they deal with information (measurement and modeling) that is
geographically distributed on (a portion of) a planetary surface. In
our own recent series of
papers~\cite[]{Wieczorek+2005,Simons+2006a,Simons+2006b,Simons+2007,Wieczorek+2007,Dahlen+2008,Simons+2009b,Plattner+2012,Plattner+2013}
we have dealt extensively with Slepian's problem in spherical
geometry; asymptotic reductions to the
plane~\cite[]{Simons+2006a,Simons+2011a} then generalize Slepian's early
treatment of the multidimensional Cartesian case~\cite[]{Slepian64}.

In this chapter we provide a framework for the analysis and
representation of geoscientific data by means of Slepian functions
defined for time series, on two-dimensional Cartesian, and spherical
domains. We emphasize the common ground underlying the construction of
all Slepian functions, discuss practical algorithms, and review the
major findings of our own recent work on
signal~\cite[]{Wieczorek+2005,Simons+2006b} and power spectral estimation
theory on the sphere~\cite[]{Wieczorek+2007,Dahlen+2008}. Compared to the
first edition of this work~\cite[]{Simons2010} we now also include a
section on \textit{vector}-valued Slepian functions that brings the
theory in line with the modern demands of (satellite) gravity,
geomagnetic or oceanographic data
analysis~\cite[]{Freeden2010,Olsen+2010,Grafarend+2010,Martinec2010,Sabaka+2010}.

\section{Theory of Slepian functions}
 
In this section we review the theory of Slepian functions in one
dimension, in the Cartesian plane, and on the surface of the unit
sphere. The one-dimensional theory is quite well known and perhaps
most accessibly presented in the textbook by Percival and
Walden~\cite[]{Percival+93}. It is briefly reformulated here for
consistency and to establish some notation. The two-dimensional planar
case formed the subject of a lesser-known of Slepian's
papers~\cite[]{Slepian64} and is reviewed here also. We are not
discussing alternatives by which two-dimensional Slepian functions are
constructed by forming the outer product of pairs of one-dimensional
functions. While this approach has produced some useful
results~\cite[]{Hanssen97,Simons+2000}, it does not solve the
concentration problem \textit{sensu stricto}. The spherical scalar
case was treated in most detail, and for the first time, by ourselves
elsewhere~\cite[]{Wieczorek+2005,Simons+2006a,Simons+2006b}, though two
very important early studies by Slepian, Gr\"unbaum, and others, laid
much of the foundation for the analytical treatment of the spherical
concentration problem for cases with special symmetries
~\cite[]{Gilbert+77,Grunbaum81a}. The spherical vector case was treated
in its most general form by ourselves
elsewhere~\cite[]{Plattner+2012,Plattner+2013}, but had also been studied
in some, but less general, detail by researchers interested in medical
imaging~\cite[]{Maniar+2005,Mitra+2006} and
optics~\cite[]{Jahn+2012,Jahn+2013a}. Finally, we recast the theory in
the context of reproducing-kernel Hilbert spaces, through which the
reader may appreciate some of the connections with radial basis
functions, splines, and wavelet analysis, which are commonly
formulated in such a framework~\cite[]{Freeden+98c,Michel2010}.

\ssec{Spatiospectral Concentration for Time Series}

\sssec{General Theory in One Dimension}

We use~$t$ to denote time or one-dimensional space and~$\omega$ for
angular frequency, and adopt a normalization convention~\cite[]{Mallat98}
in which a real-valued time-domain signal~$f(t)$ and its Fourier
transform~$F(\omega)$ are related by 
\be
\label{slepian0}
f(t)=\jnorm\intinf 
F(\omega)e^{i\omega t}\,d\omega,\qquad 
F(\omega)=\intinf 
f(t)e^{-i\omega t}\,dt.
\ee
The problem of finding the strictly bandlimited signal 
\be
\label{gband}
g(t)=\jnorm\intW 
G(\omega)e^{i\omega t}\,d\omega
,
\ee
that is maximally, though by virtue of the Paley-Wiener
theorem~\cite[]{Daubechies92,Mallat98} never completely, concentrated
into a time interval $|t|\le T$ was first considered by Slepian,
Landau and Pollak~\cite[]{Slepian+61,Landau+61}. The optimally
concentrated signal is taken to be the one with the least energy
outside of the interval: 
\be
\lambda=\fracd{\int_{-T}^{T}g^2(t)\,dt}
{\intinf g^2(t)\,dt}=\mbox{maximum}.
\label{slepian1}
\ee
Bandlimited functions~$g(t)$ satisfying the variational
problem~(\ref{slepian1}) have spectra~$G(\omega)$ that satisfy the
frequency-domain convolutional integral eigenvalue equation
\begin{subequations}
\label{slepian2}
\be
\intW D(\omega,\omega')\,G(\omega')\,d\omega'
=\lambda\hspace{0.05em}G(\omega),\quad |\omega|\le W,
\ee
\be
D(\omega,\omega')=\frac{\sin T(\omega-\omega')}
{\pi (\omega-\omega')}.
\ee
\end{subequations}
The corresponding time- or spatial-domain formulation is
\begin{subequations}
\label{dude1}
\be
\label{eig1}
\int_{-T}^{T}\Dttp\,g(t')\,dt'=\lambda\hspace{0.05em}g(t),\quad
t\in\mathbb{R}
,
\ee
\be
\label{slepian4}
\Dttp=\frac{\sin W(t-t')}{\pi (t-t')}
.
\ee
\end{subequations}
The ``prolate spheroidal eigentapers'' $g_1(t),g_2(t),\ldots$ that
solve eq.~(\ref{dude1}) form a
doubly orthogonal set. When they are chosen to be orthonormal over 
infinite time $|t|\le\infty$ they are also orthogonal over
the finite interval $|t|\le T$: 
\be
\intinf g_{\alpha}g_{\beta}\,dt=\delta_{\alpha\beta}
,\qquad\int_{-T}^{T}g_{\alpha}g_{\beta}\,dt=
\lambda_{\alpha}\hspace{0.05em}\delta_{\alpha\beta}.
\label{slepian8}
\ee
A change of variables and a scaling of the eigenfunctions transforms
eq.~(\ref{slepian2}) into the dimensionless eigenproblem
\begin{subequations}
\label{slepian5}
\be
\int_{-1}^{1}D(x,x')\,\psi(x')\,dx'=\lambda\hspace{0.05em}\psi(x),
\ee
\be
D(x,x')=\fracd{\sin TW(x-x')}{\pi (x-x')}.
\ee
\end{subequations}
Eq.~(\ref{slepian5}) shows that the eigenvalues 
$\lambda_1>\lambda_2>\ldots$ and suitably scaled eigenfunctions
$\psi_1(x),\psi_2(x),\ldots$ depend only upon the time-bandwidth
product~$TW$. The sum of the concentration eigenvalues~$\lambda$ 
relates to this product by  
\be
N\D=\sum_{\alpha=1}^{\infty}\lambda_{\alpha}
=\int_{-1}^{1}D(x,x)\,dx
=\frac{(2T)(2W)}{2\pi}= 
\frac{2TW}{\pi}. 
\label{slepian6}
\ee
The eigenvalue spectrum of eq.~(\ref{slepian5}) has a
characteristic step shape, showing significant ($\lambda\approx 1$)
and insignificant ($\lambda\approx 0$) eigenvalues separated by a
narrow transition band~\cite[]{Landau65,Slepian+65}. Thus, this
``Shannon number'' is a good estimate of the number of
significant eigenvalues, or, roughly speaking, $N\D$ is the number of signals
$f(t)$ that can be simultaneously well concentrated into a finite time
interval $|t|\le T$ and a finite frequency interval $|\omega|\le
W$. In other words~\cite[]{Landau+62}, $N\D$ is the approximate dimension
of the space of signals that is ``essentially'' timelimited to~$T$ and
bandlimited to~$W$, and using the orthogonal set
$g_1,g_2,\ldots,g_{N\D}$ as its basis is parsimonious. 

\sssec{Sturm-Liouville Character and Tridiagonal Matrix Formulation}

The integral operator acting upon~$\psi$ in eq.~(\ref{slepian5})
commutes with a differential operator that arises in expressing the
three-dimensional scalar wave equation in prolate spheroidal
coordinates~\cite[]{Slepian+61,Slepian83}, which makes it possible to
find the scaled eigenfunctions~$\psi$ by solving the
Sturm-Liouville equation  
\be
\label{slepian7.5}
\frac{d}{dx}\left[(1-x^2)\frac{d\psi}{dx}\right]+
\left[\chi-\frac{(N\D)^2\pi^2}{4}x^2\right]  
\hsp\psi=0, 
\ee
where~$\chi\ne\lambda$ is the associated eigenvalue. The
eigenfunctions~$\psi(x)$ of eq.~(\ref{slepian7.5}) can be found at
discrete values of~$x$ by diagonalization of a simple symmetric
tridiagonal matrix~\cite[]{Slepian78,Grunbaum81a,Percival+93} with
elements % VERIFIED WITH PERCIVAL'S OWN FUNCTION, EVEN, AND SEE ALSO
	 % DELSARTE 85 FOR EXAMPLE.
\ber % FJS say that this is for N samples
([N-1-2x]/2)^2\cos(2\pi W)\for x=0,\cdots,N-1,\nnr\\
x(N-x)/2\for x=1,\ldots,N-1
.
\eer
The matching eigenvalues~$\lambda$ can then be obtained directly from
eq.~(\ref{slepian5}). The discovery of the Sturm-Liouville formulation
of the concentration problem posed in eq.~(\ref{slepian1}) proved to
be a major impetus for the widespread adoption and practical
applications of the ``Slepian'' basis in signal identification,
spectral analysis and numerical analysis. Compared to the sequence of
eigenvalues~$\lambda$, the spectrum of the eigenvalues~$\chi$ is
extremely regular and thus the solution of eq.~(\ref{slepian7.5}) is
without any problem amenable to finite-precision numerical
computation~\cite[]{Percival+93}.

\ssec{Spatiospectral Concentration in the Cartesian Plane}

\sssec{General Theory in Two Dimensions}

A square-integrable function~$f(\bx)$ defined in
the plane has the two-dimensional Fourier representation 
\be
\label{Bfourier}
f(\bx)=\fnorm\intinf
F(\bk)e^{i\bk\cdot\bx}\dbk,\qquad
F(\bk)=\intinf
f(\bx)e^{-i\bk\cdot\bx}\dbx,
\ee
We use~$g(\bx)$ to denote a function that is bandlimited
to~$\mathcal{K}$, an arbitrary subregion of spectral space, 
\be
\label{Bgdefn}
g(\bx)=\fnorm\!\!
\intK G(\bk)e^{i\mbf{k}\cdot\bx}\dbk
.
\ee
Following \cite{Slepian64}, we seek to concentrate the power
of~$g(\bx)$ into a finite spatial region~$\mathcal{R}\in\mathbb{R}^2$,
of area~$A$: 
\be
\label{Bnormratio}
\lambda=\fracd{\intR g^2(\bx)\dbx}
{\intinf g^2(\bx)\dbx}=\mbox{maximum}.
\ee
Bandlimited functions~$g(\bx)$ that maximize the Rayleigh
quotient~(\ref{Bnormratio}) solve the Fredholm integral
equation~~\cite[]{Tricomi70} 
\begin{subequations}
\label{Beigen12}
\be
\label{Beigen1}
\intK \Dkkp\,G(\bk')\dbk'
=\lambda\hsp G(\bk),\qquad\bk\in\mathcal{K}
,
\ee
\be
\label{Beigen2}
\Dkkp=\fnorm
\intR e^{i(\bk'-\bk)\cdot\bx}\dbx
.
\ee
\end{subequations}
The corresponding problem in the spatial domain is  
\begin{subequations}
\label{Beigen34}
\be
\label{eig2}
\intR\! \Dxxp\,g(\bx')\dbx'
=\lambda\hsp g(\bx),\qquad\bx\in\mathbb{R}^2,
\ee
\be
\label{Beigen4}
\Dxxp=\fnorm
\intK e^{i\bk\cdot(\bx-\bx')}\dbk.
\ee
\end{subequations}
The bandlimited spatial-domain eigenfunctions
$g_1(\bx),g_2(\bx),\ldots$ and eigenvalues $\lambda_1\ge\lambda_2\ge\ldots$
that solve eq.~(\ref{Beigen34}) may be chosen to be orthonormal over
the whole plane $\|\bx\|\le\infty$ in which case they are also
orthogonal over~$\mathcal{R}$: 
\be
\intinf
g_{\alpha}g_{\beta}\dbx=\delta_{\alpha\beta},\qquad
\label{Borthog1}
\intR g_{\alpha}g_{\beta}\dbx=\lambda_{\alpha}\delta_{\alpha\beta}.
%\label{Borthog2}
\ee
Concentration to the disk-shaped spectral band
$\mathcal{K}=\{\bk:\|\bk\|\le K\}$ allows us to rewrite
eq.~(\ref{Beigen34}) after a change of variables and a 
scaling of the eigenfunctions as 
\begin{subequations}
\label{Dscaled}
\be
\int_{\mathcal{R}_{*}}\!\!\Dxixip
\,\psi(\bxi')\,d\bxi'=\lambda\hsp\psi(\bxi),
\ee
\be
\Dxixip=\frac{K\sqrt{A/4\pi}}{2\pi}
\fracd{J_1(K\sqrt{A/4\pi}\,\,\|\bxi-\bxi'\|)} 
{\|\bxi-\bxi'\|},
\ee
\end{subequations}
where the region~$\mathcal{R}_*$ is scaled to area~$4\pi$ and
$J_1$ is the first-order Bessel function of the first
kind. Eq.~(\ref{Dscaled}) shows that, also in the two-dimensional
case, the eigenvalues $\lambda_1,\lambda_2,\ldots$ and the scaled
eigenfunctions $\psi_1(\bxi),\psi_2(\bxi),\ldots$ depend only on the
combination of the circular bandwidth~$K$ and the spatial
concentration area~$A$, where the quantity $K^2A/(4\pi)$ now plays the
role of the time-bandwidth product~$TW$ in the one-dimensional
case. The sum of the concentration eigenvalues~$\lambda$ defines the 
two-dimensional Shannon number~$N\DD$ as 
\be
\label{Bshannon}
N\DD=\sum_{\alpha=1}^{\infty}\lambda_{\alpha}
=\int_{R_{*}}\!
\Dxixi\,d\bxi
=\fracd{(\pi K^2)(A)}{(2\pi)^2}
= K^2\,\fracd{A}{4\pi}
.
\ee
Just as~$N\D$ in eq.~(\ref{slepian6}), $N\DD$ is the product of
the spectral and spatial areas of concentration multiplied by the
``Nyquist density''~\cite[]{Daubechies88a,Daubechies92}. And, similarly, it
is the effective dimension of the space of ``essentially'' space- and
bandlimited functions in which the set of two-dimensional functions
$g_1,g_2,\dots,g_{N\DD}$ may act as a sparse orthogonal basis.   

\begin{figure}\center
\rotatebox{-90}{\includegraphics[height=\columnwidth]{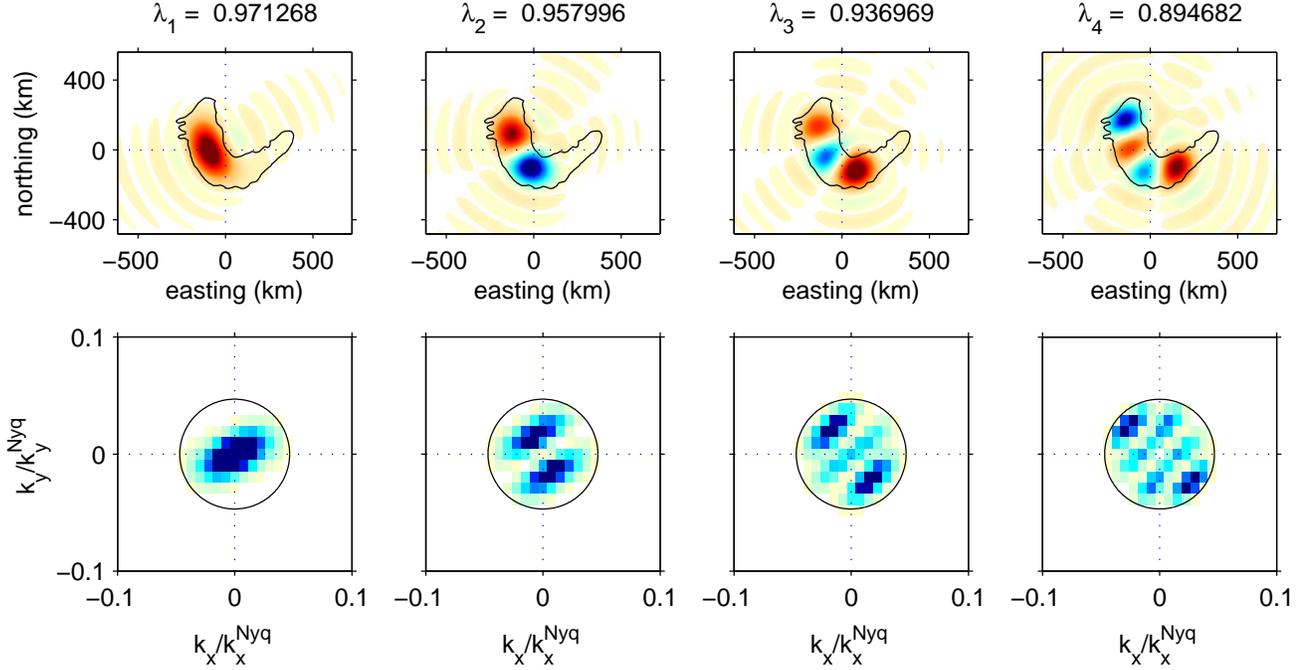}}
\caption{\label{swregions2d_2}Bandlimited eigenfunctions
$g_1,g_2,\ldots,g_{4}$ that are optimally concentrated within the
Columbia Plateau, a physiographic region in the United States centered 
on 116.02$^\circ$W 43.56$^\circ$N (near Boise City, Idaho) of area
$A\approx145\times 10^3$~km$^2$. The concentration factors
$\lambda_1,\lambda_2,\ldots,\lambda_{4}$ are indicated; the Shannon
number $N\DD=10$. The top row shows a rendition of the eigenfunctions
in space on a grid with 5~km resolution in both directions, with the
convention that positive values are blue and negative values red,
though the sign of the functions is arbitrary. The
spatial concentration region is outlined in black. The bottom row
shows the squared Fourier coefficients $|G_\alpha(\mbf{k})|^2$ as
calculated from the functions $g_\alpha(\bx)$ shown, on a wavenumber
scale that is expressed as a fraction of the Nyquist wavenumber. The
spectral limitation region is shown by the black circle at wavenumber
$K=0.0295$~rad/km. All areas for which the absolute value of the
functions plotted is less than one hundredth of the maximum value
attained over the domain are left white. The calculations were
performed by the Nystr\"om method using Gauss-Legendre integration of
eq.~(\ref{Dscaled}) in the two-dimensional spatial domain~\cite[]{Simons+2011a}.}
\end{figure}

After a long hiatus since the work of \cite{Slepian64}, the
two-dimensional problem has recently been the focus of renewed
attention in the applied mathematics
community~\cite[]{DeVilliers+2003,Shkolnisky2007}, and applications to
the geosciences are following suit~\cite[]{Simons+2011a}. Numerous
numerical methods exist to use eqs~(\ref{Beigen12})--(\ref{Beigen34})
in solving the concentration problem~(\ref{Bnormratio}) on
two-dimensional Cartesian domains. An example of Slepian functions on
a geographical domain in the Cartesian plane can be found in
Figure~\ref{swregions2d_2}.

\sssec{Sturm-Liouville Character and Tridiagonal Matrix Formulation}

If in addition to the circular spectral limitation, space is also
circularly limited, in other words, if the spatial region of
concentration or limitation~$\mathcal{R}$ is a circle of radius~$R$,
then a polar coordinate, $\bx=(r,\theta)$, representation 
\be g(r,\theta)=\left\{
\begin{array}{l@{\quad\mbox{if}\hspace{0.6em}}l}
\rule[-2mm]{0mm}{6mm}\sqrt{2}\,g(r)\cos m\theta & m<0,\\
\rule[-2mm]{0mm}{6mm}g(r)                     & m=0,\\
\rule[-2mm]{0mm}{6mm}\sqrt{2}\,g(r)\sin m\theta & m>0,\\
\end{array}
\right.
\label{Bpolarg}
\ee
may be used to decompose eq.~(\ref{Dscaled}) into a series of
non-degenerate fixed-order eigenvalue problems, after scaling,  
\begin{subequations}
\label{Dscaled2}
\be
\label{Bscaled2}
\int_{0}^{1}D(\xi,\xi')\hsp\psi(\xi')\,
\xi'\hspace{0.05em}d\xi'=\lambda\hsp\psi(\xi),
\ee
\be % FJS PROOF
D(\xi,\xi')=4N\DD\int_{0}^{1}\!
J_m\big(2\sqrt{N\DD}\,p\hspace{0.05em}\xi\big)\,
J_m\big(2\sqrt{N\DD}\,p\hspace{0.05em}\xi'\big)\,p\hsp dp
.
\ee
\end{subequations}
The solutions to eq.~(\ref{Dscaled2}) also solve a Sturm-Liouville
equation on $0\le\xi \le 1$. In terms of
$\varphi(\xi)=\sqrt{\xi}\,\psi(\xi)$, 
\be
\fracd{d}{d\xi}\left[(1-\xi^2)\frac{d\varphi}{d\xi}\right] + \left( \chi+
\fracd{1/4-m^2}{\xi^2} -4N\DD \xi^2  \right)\varphi =0
\label{gpsf},
\ee
for some~$\chi\ne\lambda$. When~$m=\pm1/2$ eq.~(\ref{gpsf}) reduces to
eq.~(\ref{slepian7.5}). By extension to~$\xi>1$
the fixed-order ``generalized prolate spheroidal functions''
$\varphi_1(\xi),\varphi_2(\xi),\dots$ can be determined from the
rapidly converging infinite series
\be
%% \varphi(\xi)=\frac{m!}{\gamma}\sum_{l=0}^\infty  
%% \frac{d_l\,l!}{(l+m)!}
%% \frac{J_{m+2l+1}(c\hsp\xi)}{\sqrt{c\hsp\xi}},\qquad \xi\in\mathbb{R}^+,
\varphi(\xi)=
\frac{\sqrt{2}}{\gamma}
\sum_{l=0}^\infty  
 (2l+m+1)^{1/2} d_l
\frac{J_{m+2l+1}(c\hsp\xi)}{\sqrt{c\hsp\xi}},\qquad \xi\in\mathbb{R}^+,
\label{SE}
\ee
where $\varphi(0)=0$ and the fixed-$m$ expansion coefficients~$d_l$ are
determined by recursion~\cite[]{Slepian64} or by diagonalization of a
symmetric tridiagonal matrix~\cite[]{DeVilliers+2003,Shkolnisky2007}
with elements given by
\begin{eqnarray}\label{Telements}
%T_{l+1\,l}&=&-\fracd{c^2\,(m+l+1)^2}{(2l+m+1)(2l+m+2)},\nnr\\
T_{ll}&=&\left( 2l + m + \frac{1}{2}\right) \left(
2l+m+\frac{3}{2} \right)+\fracd{c^2}{2} \left[1+
\fracd{m^2}{(2l+m)(2l+m+2)} \right],\nnr\\
T_{l+1\,l}&=&-\fracd{c^2\,(l+1)(m+l+1)}{\sqrt{2l+m+1}\,(2l+m+2)\,\sqrt{2l+m+3}},%=T_{l\,l+1},
%,\nnr\\
%T_{l\,l+1}&=&-\fracd{c^2\,(l+1)^2}{(2l+m+2)(2l+m+3)},
\end{eqnarray}
where the parameter~$l$ ranges from~$0$ to
some large value that ensures convergence. 
%% The symmetrized
%% version~\cite[]{Shkolnisky2007}  
%% \begin{eqnarray}\label{Telementsprime}
%% T'_{l+1\,l}&=&-\fracd{c^2\,(l+1)(m+l+1)}
%% {\sqrt{2l+m+1}\,(2l+m+2)\,\sqrt{2l+m+3}}=T'_{l\,l+1},
%% \end{eqnarray}
%% preserves the diagonal, $T'_{ll}=T_{ll}$, and has exactly the same
%% eigenvalues as~$T$.
The desired concentration eigenvalues~$\lambda$
can subsequently be obtained by direct integration of
eq.~(\ref{Dscaled}), or, alternatively, from  
\be
\lambda=2\gamma^2\sqrt{N\DD},\with\gamma=
\frac{c^{m+1/2}d_0}{2^{m+1}(m+1)!}
\left(\sum_{l=0}^\infty d_l\right)^{-1} 
.
\ee

\begin{figure}\center
\includegraphics[width=0.9\columnwidth]{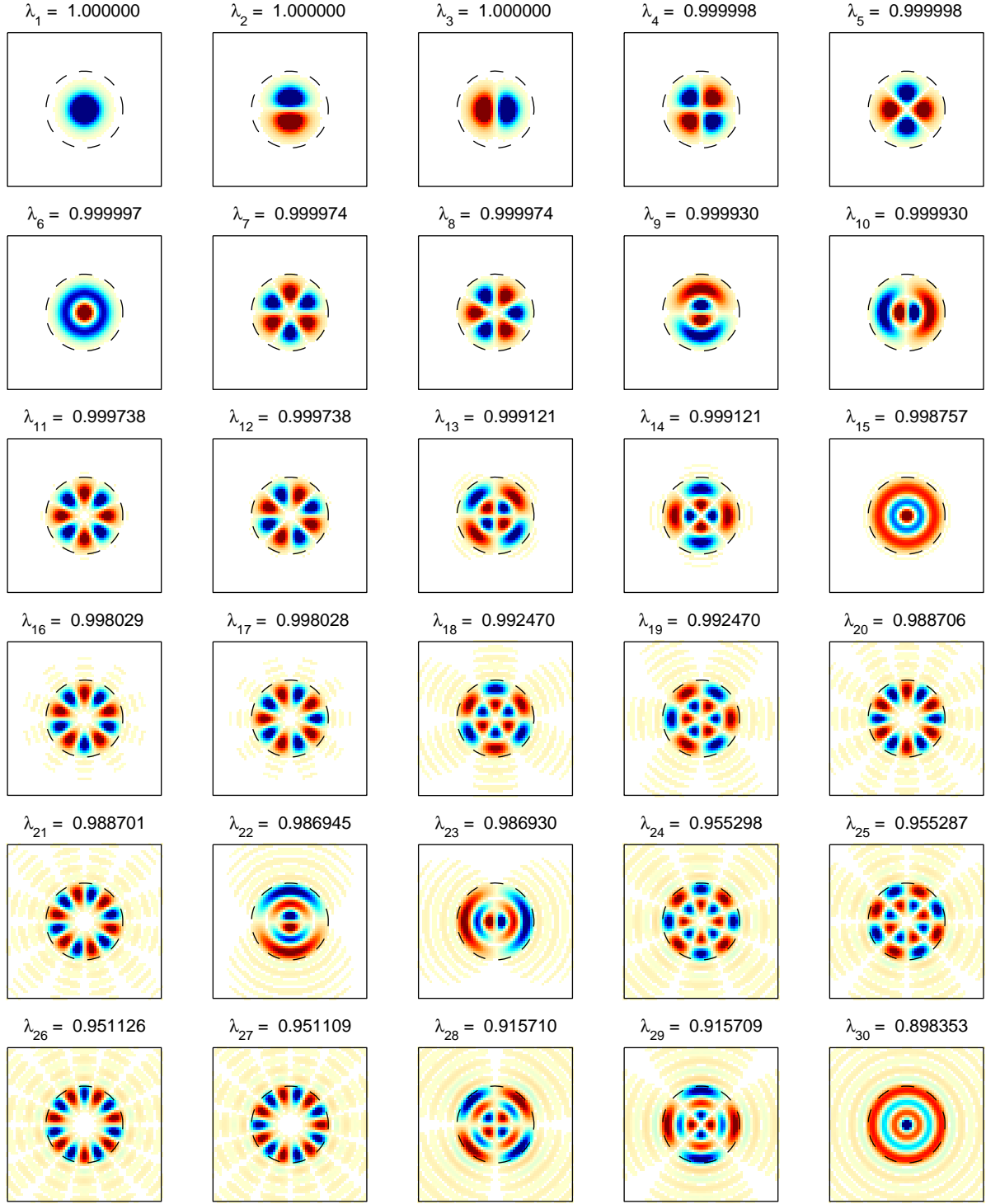}
\caption{\label{swfried2d}Bandlimited eigenfunctions~$g_\alpha(r,\theta)$
that are optimally concentrated within a Cartesian disk of
radius~$R=1$. The dashed circle denotes the region boundary. The Shannon
number~$N\DD=42$. The eigenvalues $\lambda_{\alpha}$ have been
sorted to a global ranking with the best concentrated eigenfunction
plotted at the top left and the $30$th best in the lower right. Blue
is positive and red is negative and the color axis is symmetric, but
the sign is arbitrary; regions in which the absolute value is less
than one hundredth of the maximum value on the domain are left
white. The calculations were performed by Gauss-Legendre integration
in the two-dimensional spatial domain, which sometimes leads to slight
differences in the last two digits of what should be identical
eigenvalues for each pair of non-circularly-symmetric eigenfunctions.}
\end{figure}

An example of Slepian functions on a disk-shaped region in the
Cartesian plane can be found in Figure~\ref{swfried2d}. The solutions
were obtained using the Nystr\"om method using Gauss-Legendre
integration of eq.~(\ref{Dscaled}) in the two-dimensional spatial
domain. These differ only very slightly from the results of
computations carried out using the diagonalization of
eqs~(\ref{Telements}) directly, as shown and discussed by us
elsewhere~\cite[]{Simons+2011a}. 

\ssec{Spatiospectral Concentration on the Surface of a Sphere}

\sssec{General Theory in ``Three'' Dimensions}

We denote the colatitude of a geographical point~$\rhat$ on the
unit sphere surface $\Omega=\{\rhat: \|\rhat\|=1\}$ by
$0\le\theta\le\pi$ and the longitude by $0\le\phi< 2\pi$. We use~$R$
to denote a region of~$\Omega$, of area~$A$, within which we seek to
concentrate a bandlimited function of position~$\rhat$. We express a
square-integrable function~$f(\rhat)$ on the surface of the unit
sphere as 
\be
\label{expansion}
f(\rhat)=\sumsh {f}_{lm}\Ylmrh,\qquad {f}_{lm}=\into
f(\rhat)\hspace*{0.05em}\Ylmrh\domg,
\ee
using orthonormalized real surface spherical
harmonics~~\cite[]{Edmonds96,Dahlen+98} 
\ber
\Ylmrh= \Ylm(\theta,\phi)&=&\left\{
\begin{array}{l@{\quad\mbox{if}\hspace{0.6em}}l}
\rule[-2mm]{0mm}{6mm}\sqrt{2}\Xlamth\cos m\phi & -l\le m<0,\\
\rule[-2mm]{0mm}{6mm}X_{l0}(\theta)                     & m=0,\\
\rule[-2mm]{0mm}{6mm}\sqrt{2}\Xlm(\theta)\sin m\phi & 0< m\le l,\\
\end{array}
\right.\label{Ylm}\\
\Xlm(\theta)&=&
(-1)^m\tlofp^{1/2}
\left[\frac{(l-m)!}{(l+m)!}\right]^{1/2}\!\Plm (\cos\theta),
\label{xlm}\\
\Plm (\mu)&=&
\frac{1}{2^ll!}(1-\mu^2)^{m/2}\left(\frac{d}{d\mu}\right)^{l+m}\!(\mu^2-1)^l.
\label{plm}
\eer
The quantity $0\le l\le\infty$ is the angular degree of the
spherical harmonic, and $-l\le m\le l$ is its angular order. The
function~$\Plm(\mu)$ defined in (\ref{plm}) is the associated
Legendre function of integer degree~$l$ and order~$m$. Our
choice of the constants in eqs~(\ref{Ylm})--(\ref{xlm})
orthonormalizes the harmonics on the unit sphere:
\be
\into\Ylm\Ylmp\domg=\dllp\dmmp
,
\label{normalization}
\ee
and leads to the addition theorem in terms of the Legendre
functions~$P_l(\mu)=P_{l0}(\mu)$ as
\be
\suml_{m=-l}^l\Ylmrh\Ylmrhp =
\tlofp P_l(\rhat\cdot\rhat').
\label{additionSH}
\ee
To maximize the spatial concentration of a bandlimited
function
\be
g(\rhat)=\sumshL g_{lm}\Ylmrh
\label{bandlg}
\ee
within a region~$R$, we maximize the energy ratio
\be
\lambda=\fracd{\intr g^2(\rhat)\domg}{\into^{}g^2(\rhat)\domg}=\mbox{maximum}
.
\label{normratio}
\ee
Maximizing equation~(\ref{normratio}) leads to the 
positive-definite spectral-domain eigenvalue equation
\begin{subequations}
\label{fulleigen1}
\be
\label{halfeigen}
\suml_{l'=0}^L\suml_{m'=-l'}^{l'}\Dlmlmp
g_{l'm'}=\lambda\hspace{0.05em}g_{lm},\qquad 0\le l\le L,
\ee
\be
\Dlmlmp=\intr\Ylm\Ylmp\domg,
\label{Dlmlmpdef}
\ee
\end{subequations}
and we may equally well rewrite eq.~(\ref{fulleigen1}) as a
spatial-domain eigenvalue equation: 
\begin{subequations}
\label{firsttimeint}
\be
\label{eig3}
\intr  \Drhrhp \,\grhp\domg'=
\lambda\hspace{0.05em}g(\rhat),\qquad \rhat\in\Omega,
\ee
\be
\Drhrhp
=\sum_{l=0}^L\tlofp\!P_l(\rhat\cdot\rhat'),
\label{banddelta}
\ee
\end{subequations}
where~$P_l$ is the Legendre function of
degree~$l$. Eq.~(\ref{firsttimeint}) is a homogeneous Fredholm
integral equation of the second kind, with a finite-rank, symmetric,
Hermitian kernel. We choose the spectral eigenfunctions of the operator in
eq.~(\ref{Dlmlmpdef}), whose elements are $\glma$, $\alpha=1,$
$\dots,$ $\Lpot$, to satisfy the orthonormality relations 
\be
\label{orthogn}
\sumshortL\galm\glmb=\dab,\qquad
\sumshortL\galm\sumshortLp\Dlmlmp\gblmp=\lambda_\alpha\dab
.
\ee
The finite set of bandlimited spatial eigensolutions
$g_1(\rhat),g_2(\rhat), \ldots,g_{(L+1)^2}(\rhat)$ can be made
orthonormal over the whole sphere~$\Omega$ and orthogonal over the
region~$R$:  
\be
\into g_{\alpha}g_{\beta}\domg=\delta_{\alpha\beta},\qquad
\intr  g_{\alpha}g_{\beta}\domg=\lambda_{\alpha}\delta_{\alpha\beta}.
\label{orthog}
\ee
In the limit of a small area $A\rightarrow 0$ and a large bandwidth 
$L\rightarrow\infty$ and after a change of variables, a scaled version
of eq.~(\ref{firsttimeint}) will be given by 
\begin{subequations}
\label{3Dscaled}
\be
\int_{R_{*}}\!\!\Dxixip
\,\psi(\bxi')\,d\Omega'_*=\lambda\hsp\psi(\bxi),
\ee
\be
\Dxixip=\frac{(L+1)\sqrt{A/4\pi}}{2\pi}
\fracd{J_1[(L+1)\sqrt{A/4\pi}\,\,\|\bxi-\bxi'\|]} 
{\|\bxi-\bxi'\|},
\ee
\end{subequations}
where the scaled region~$R_{*}$  now has area~$4\pi$ and~$J_1$ again
is the first-order Bessel function of the first kind. As in the
one- and two-dimensional case, the asymptotic, or ``flat-Earth''
eigenvalues $\lambda_1\ge\lambda_2\ge\ldots$ and scaled eigenfunctions
$\psi_1(\bxi),\psi_2(\bxi),\ldots$ depend upon the maximal
degree~$L$ and the area~$A$ only through what is once again a
space-bandwidth product, the ``spherical Shannon number'',
this time given by  
\ber
\nnr
N\DDD&=&\sum_{\alpha =1}^{\Lpot}\lambda_{\alpha}=
\sum_{l=0}^L\summ
D_{lm,lm}=\int_RD(\rhat,\rhat)\,d\Omega\\
&=&\int_{R_{*}}\!\!\Dxixi
\,d\Omega_*
=\Lpot\,\frac{A}{4\pi}
\label{tracedef}
.
\eer
Irrespectively of the particular region of concentration that they
were designed for, the complete set of bandlimited spatial Slepian
eigenfunctions $g_1,g_2, \ldots,g_{(L+1)^2}$ is a basis for
bandlimited scalar processes anywhere on the surface of the unit
sphere~\cite[]{Simons+2006a,Simons+2006b}. This follows directly from the
fact that the spectral localization kernel~(\ref{Dlmlmpdef}) is real,
symmetric, and positive definite: its eigenvectors $g_{1\hsp lm},
g_{2\hsp lm}, \ldots, g_{\Lpot \hsp lm}$ form an orthogonal set as we
have seen. Thus the Slepian basis functions $\garh$,
$\alpha=1,\ldots,\Lpot$ given by eq.~(\ref{bandlg}) simply transform
the same-sized limited set of spherical harmonics~$\Ylmrh$, $0\le l\le
L$, $-l\le m\le l$ that are a basis for the same space of bandlimited
spherical functions with no power above the bandwidth~$L$. The effect
of this transformation is to order the resulting basis set such that
the energy of the first~$N\DDD$ functions, $g_{1}(\rhat),\ldots,
g_{N\DDD}(\rhat)$, with eigenvalues $\lambda\approx 1$, is
concentrated in the region~$R$, whereas the remaining eigenfunctions,
$g_{N\DDD+1}(\rhat),\ldots,g_{\Lpot}(\rhat)$, are concentrated in the
complimentary region~$\Omega\setminus R$. As in the one- and two-dimensional
case, therefore, the reduced set of basis functions
$g_1,g_2,\ldots,g_{N\DDD}$ can be regarded as a sparse, global, basis
suitable to approximate bandlimited processes that are primarily
localized to the region~$R$. The dimensionality reduction is dependent
on the fractional area of the region of interest. In other words, the
full dimension of the space~$\Lpot$ can be ``sparsified'' to an
effective dimension of $N\DDD=\Lpot A/(4\pi)$ when the signal of
interest resides in a particular geographic region.

\begin{figure}\center
\rotatebox{-90}{\includegraphics[height=0.8\columnwidth]
{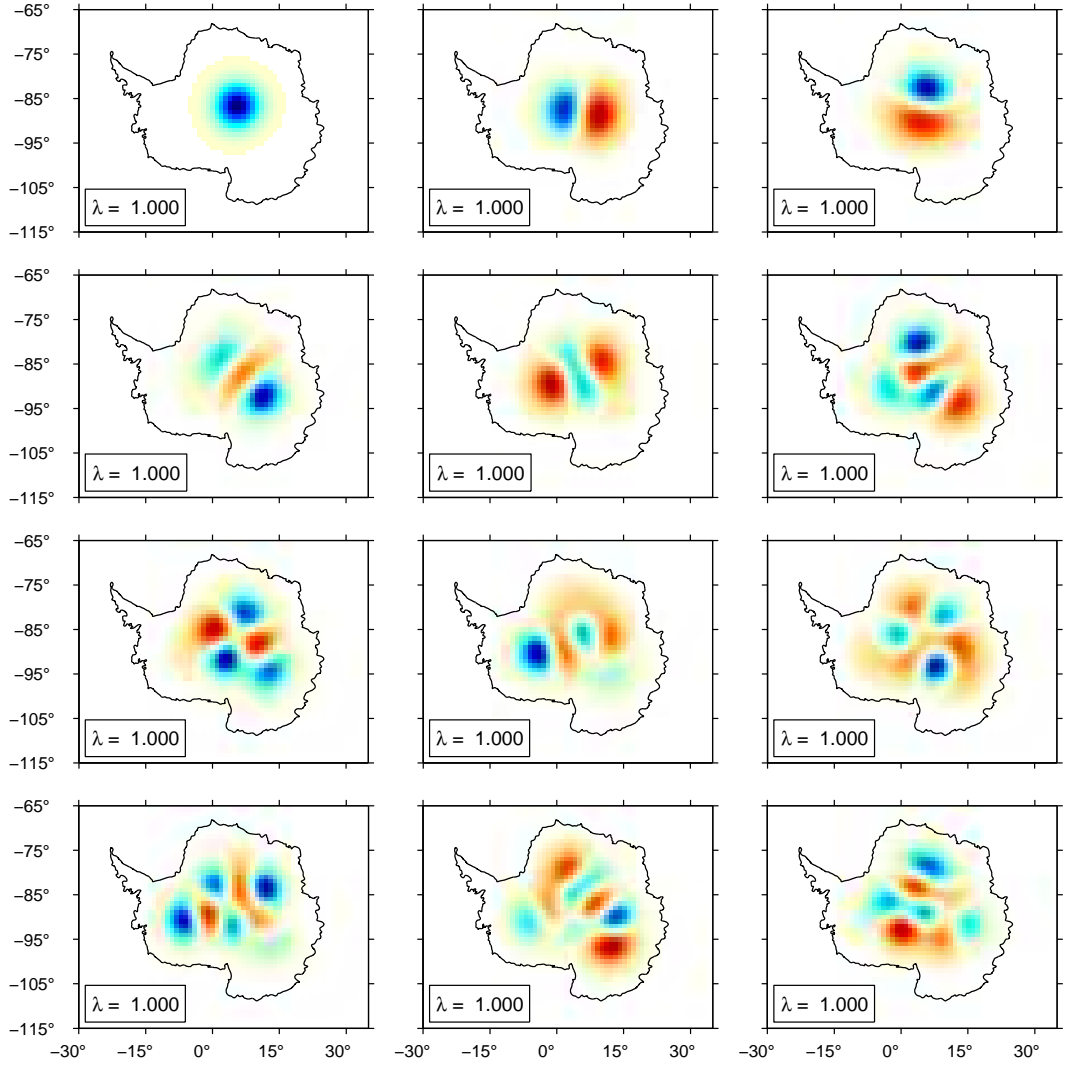}}
\caption{\label{antarctica}
Bandlimited $L=60$ eigenfunctions
$g_1,g_2,\ldots,g_{12}$ that are optimally concentrated within
Antarctica. The concentration factors
$\lambda_1,\lambda_2,\ldots,\lambda_{12}$ are indicated; the rounded
Shannon number is $N\DDD=102$.  The order of concentration is left to right,
top to bottom. Positive values are blue and negative values are red;
the sign of an eigenfunction is arbitrary. Regions in which the
absolute value is less than one hundredth of the maximum value on the
sphere are left white. We integrated eq.~(\ref{Dlmlmpdef}) over a
splined high-resolution representation of the boundary, using
Gauss-Legendre quadrature over the colatitudes, and analytically in
the longitudinal dimension~\cite[]{Simons+2007}. 
}
\end{figure}

Numerical methods for the solution of
eqs~(\ref{fulleigen1})--(\ref{firsttimeint}) on completely general
domains on the surface of the sphere were discussed by us
elsewhere~\cite[]{Simons+2006a,Simons+2006b,Simons+2007}. An example of
Slepian functions on a geographical domain on the surface of the
sphere is found in Figure~\ref{antarctica}.

\sssec{Sturm-Liouville Character and Tridiagonal Matrix Formulation}

In the special but important case in which the region of concentration
is a circularly symmetric cap of colatitudinal radius~$\Theta$,
centered on the North Pole, the colatitudinal parts~$g(\theta)$ of the
separable functions  
\be
g(\theta,\phi)=\left\{
\barray{l@{\quad\mbox{if}\hspace{0.6em}}l}
\rule[-2mm]{0mm}{6mm}\sqrt{2}\,g(\theta)\cos m\phi & -L\le m<0,\\
\rule[-2mm]{0mm}{6mm}g(\theta)                     & m=0,\\
\rule[-2mm]{0mm}{6mm}\sqrt{2}\,g(\theta)\sin m\phi & 0< m\le L,\\
\earray
\right.
\label{polarg2}
\ee
which solve eq.~(\ref{firsttimeint}), or, indeed, the fixed-order
versions 
\begin{subequations}
\be
\int_{0}^{\Theta}D(\theta,\theta')\,g(\theta')\sin\theta'\,d\theta'
={\lambda}\hspace{0.05em}g(\theta),\quad 0\le\theta\le\pi,
\label{Fredholm}
\ee
\be
D(\theta,\theta')=2\pi\suml_{l=m}^{L}\Xlmth \Xlmthp,
\label{dlx}
\ee
\end{subequations}
are identical to those of a Sturm-Liouville equation for
the~$g(\theta)$. In terms of $\mu=\cos\theta$, 
\be
\fracd{d}{d\mu}\left[(\mu-\cos\Theta)
(1-\mu^2)\fracd{dg}{d\mu}\right]+
\left(\chi+
L(L+2)\mu
-\fracd{m^2(\mu-\cos\Theta)}{1-\mu^2}
\right)g=0,
\label{grundiffeqn}
\ee
with~$\chi\ne\lambda$. This equation can be solved in the spectral
domain by diagonalization of a simple symmetric tridiagonal matrix
with a very well-behaved spectrum~\cite[]{Simons+2006a,Simons+2007}. This
matrix, whose eigenfunctions correspond to the~$g_{lm}$ of 
eq.~(\ref{bandlg}) at  constant~$m$ is given by
\ber T_{ll}&=&-l(l+1)\cos{\Theta},\nnr\\
T_{l\,l+1}&=&\big[l(l+2)-L(L+2)\big]
\sqrt{\fracd{(l+1)^2-m^2}{(2l+1)(2l+3)}}.
\label{gdefi1}
\eer
Moreover, when the region of concentration is a \textit{pair} of axisymmetric
polar caps of common colatitudinal radius~$\Theta$ centered on the
North and South Pole, the~$g(\theta)$ can be obtained by
solving the Sturm-Liouville equation
\be
\fracd{d}{d\mu}\left[(\mu^2-\cos^2\Theta)
(1-\mu^2)\fracd{dg}{d\mu}\right]
+\left(
\chi + L_p(L_p+3)\mu^2
-\fracd{m^2(\mu^2-\cos^2\Theta)}{1-\mu^2}
\right)g=0,\label{grundiffeqn2}
\ee
where~$L_p=L$ or~$L_p=L-1$ depending whether the order~$m$ of the
functions~$g(\theta,\phi)$ in eq.~(\ref{polarg2}) is odd or
even, and whether the bandwidth~$L$ itself is odd or even. In their
spectral form the coefficients of the optimally concentrated antipodal
polar-cap eigenfunctions only require the numerical diagonalization of
a symmetric tridiagonal matrix with analytically prescribed elements
and a spectrum of eigenvalues that is guaranteed to be
simple~\cite[]{Simons+2006b,Simons+2007}, namely 
\ber
T^p_{ll}&=&-l(l+1)\cos^2{\Theta}
+\frac{2}{2l+3}\left[(l+1)^2-m^2\right]\nnr\\
&&{}+[(l-2)(l+1)-L_p(L_p+3)]
\left[\frac{1}{3}-\frac{2}{3}\,
\fracd{3m^2-l(l+1)}{(2l+3)(2l-1)}\right],\nnr\\ 
T^p_{l\,l+2}&=&\fracd{\big[l(l+3)-L_p(L_p+3)\big]}{2l+3}
\sqrt{\fracd{\left[(l+2)^2-m^2\right]
\left[(l+1)^2-m^2\right]}{(2l+5)(2l+1)}}.
\label{gdefi}
\eer
The concentration values~$\lambda$, in turn, can be determined from
the defining equations~(\ref{fulleigen1}) or~(\ref{firsttimeint}). The
spectra of the eigenvalues~$\chi$ of eqs~(\ref{gdefi1})
and~(\ref{gdefi}) display roughly equant spacing, without the
numerically troublesome plateaus of nearly equal values that
characterizes the eigenvalues~$\lambda$. Thus, for the special cases
of symmetric single and double polar caps, the concentration problem
posed in eq.~(\ref{normratio}) is not only numerically feasible also
in circumstances where direct solution methods are bound to
fail~\cite[]{Albertella+99}, but essentially trivial in every
situation.  In practical applications, the eigenfunctions that are
optimally concentrated within a polar cap can be rotated to an
arbitrarily positioned circular cap on the unit sphere using standard
spherical harmonic rotation formulae
~\cite[]{Blanco+97,Dahlen+98,Edmonds96}. 

\ssec{Vectorial Slepian Functions on the Surface of a Sphere}

\sssec{General Theory in ``Three'' Vectorial Dimensions}

The expansion of a real-valued square-integrable vector
field~$\bef(\rhat)$ on the unit sphere~$\Omega$ can be written as
\begin{subequations}
\be
\label{general representation}
 \bef(\rhat)= \sumsh \flmP\bPlm(\rhat)+ \flmB\bBlm(\rhat) + \flmC\bClm(\rhat),
\ee
\be\label{UlmVlmWlm}
 \flmP=\into \bPlm(\rhat) \cdot \bef(\rhat)\domg,\quad
 \flmB=\into \bBlm(\rhat) \cdot \bef(\rhat)\domg,\quad\mbox{and}\,\,
 \flmC=\into \bClm(\rhat) \cdot \bef(\rhat)\domg,
\ee
\end{subequations}
using real vector surface spherical harmonics~\cite[]{Dahlen+98,Sabaka+2010,Gerhards2011} that
are constructed from the scalar harmonics in eq.~(\ref{Ylm}), as
follows. In the vector spherical coordinates $(\rhat,\thvec,\phvec)$
and using the surface gradient $\bnabla_1 = \thvec\hsom\plth +
\phvec \hsom\divsin \plphi$, we write for $l>0$ and $-m\leq l \leq m$, 
\ber
\bPlm(\rhat)& =&\rhat \Ylm(\rhat),\label{Plm}\\
\bBlm(\rhat)& =&\frac{ \bnabla_1 \Ylm(\rhat)}{\renormsing} = \frac{[
  \thvec\hsom\plth + \phvec\hsom\divsin \plphi]
  \Ylm(\rhat)}{\renormsing},\label{Blm}\\ 
\bClm(\rhat)& =&\frac{-\rhat\times \bnabla_1
  \Ylm(\rhat)}{\renormsing}=\frac{[\thvec\hsom\divsin \plphi-\phvec\hsom\plth]
  \Ylm(\rhat)}{\renormsing}, \label{Clm}
\eer
together with the purely radial $\Poo=(4\pi)^{-1/2}\hsom\rhat$, and
with $\flmB=\flmC=0$. The remaining expansion
coefficients~(\ref{UlmVlmWlm}) are naturally obtained from
eq.~(\ref{general representation}) through the orthonormality relationships 
\begin{subequations}
\ber
\label{orthonormality}
 \into \bPlm\cdot\bPlmp\domg&=&\into
 \bBlm\cdot\bBlmp\domg=\into
 \bClm\cdot\bClmp\domg=\dllp\dmmp,\\ 
\label{orthonormality 2} 
 \into \bPlm\cdot\bBlmp\domg&=&\into
 \bPlm\cdot\bClmp\domg=\into \bBlm\cdot\bClmp\domg=0 
.
\eer
\end{subequations}
The vector spherical-harmonic addition theorem~~\cite[]{Freeden+2009}
implies the limited result
\be
\label{vecadd1}
\summ\bPlm(\rhat)\cdot \bPlm(\rhat)=
\left(\frac{2l+1}{4\pi}\right)
=\summ\bBlm(\rhat)\cdot \bBlm(\rhat)=
\summ\bClm(\rhat)\cdot \bClm(\rhat).
\ee
As before we seek to maximize the spatial concentration of a
bandlimited spherical vector function
\be
\label{bandlimited field}
\bg(\rhat) = \sumshL \glmP\bPlm(\rhat) + \glmB\bBlm(\rhat) +
\glmC\bClm(\rhat) 
\ee
within a certain region $R$, in the vectorial case by maximizing the
energy ratio
\be
\label{spatial concentration equation}
 \lambda = \frac{\displaystyle
   \intr\bg\cdot\bg\domg}{\displaystyle \into
   \bg\cdot\bg\domg}=\text{maximum}. 
\ee
The maximization of eq.~(\ref{spatial concentration equation}) leads
to a coupled system of positive-definite spectral domain eigenvalue
equations 
\begin{subequations}
\label{altogether}
\ber
 \sumshLp \Dlmlmp \glmpP&=&
\lambda \hspace{0.05em}\glmP,\label{radial matrix} 
\qquad 0\le l\le L,\\
 \sumshLp \Blmlmp \glmpB + \Clmlmp \glmpC&=&
\lambda\hspace{0.05em}
 \glmB,\label{tangential matrix V}
\qquad 0\le l\le L,\\ 
 \sumshLp \Clmlmp\Tit \glmpB + \Blmlmp \glmpC&=&
\lambda\hspace{0.05em}
 \glmC,
\qquad 0\le l\le L.\label{tangential matrix W} 
\eer
\end{subequations}
Of the below matrix elements that complement the equations above,
eq.~(\ref{Dlmlmpagain}) is identical to eq~(\ref{Dlmlmpdef}),
\begin{subequations}
\label{collectlater}
\ber
\Dlmlmp&=& \intr \bPlm\cdot\bPlmp\domg=\intr\Ylm\Ylmp\domg,\label{Dlmlmpagain}\\
\Blmlmp&=&\intr \bBlm\cdot\bBlmp\domg
=\intR \bClm\cdot\bClmp\domg,\label{Blmlmp elements}\\
\Clmlmp&=&\intr \bBlm\cdot\bClmp\domg,\label{Clmlmp elements}
\eer
\end{subequations}
and the transpose of eq.~(\ref{Clmlmp elements}) switches its sign.
The radial vectorial concentration problem~(\ref{radial
  matrix})--(\ref{Dlmlmpagain}) is identical to the corresponding
scalar case~(\ref{Dlmlmpdef}), and can be solved separately from the
tangential equations. Altogether, in the space domain, the equivalent
eigenvalue equation is 
\begin{subequations}\label{nofurther}
\be\label{nofurthera}
 \intr \bDrhrhp\cdot  \bg(\rhat') \domg = \lambda\hsom\bg(\rhat), \quad
 \rhat\in \Omega,
\ee
\be\label{bandlimited Dirac} 
\bDrhrhp=\sumshL\bPlm(\rhat)\bPlm(\rhat')+\bBlm(\rhat)\bBlm(\rhat')+
\bClm(\rhat)\bClm(\rhat'), 
\ee
\end{subequations}
a homogeneous Fredholm integral equation with a finite-rank,
symmetric, separable, bandlimited kernel. Further reducing
eq.~(\ref{nofurther}) using the full version of the vectorial addition
theorem does not yield much additional insight. 
% THE OBVIOUS REDUCTION USING THE FULL ADDITION THEOREM IS NOT AT ALL
% VERY SIMPLE, SEE FREEDEN+2009, SO WE LEAVE IT AT THIS HERE. BUT IF
% WE DIDN'T WE WOULD NEED HIS PAGE 242 THEOREM 5.3.4.

After collecting the spheroidal (radial, consoidal) and toroidal
expansion coefficients in a vector,
\be
\beg=(\ldots,\glmP,\ldots,\glmB,\ldots,\glmC,\ldots)\T
\ee
and the kernel elements $\Dlmlmp$, $\Blmlmp$ and $\Clmlmp$ of
eq.~(\ref{collectlater}) into the
submatrices $\sP$,  $\sB$, and $\sC$, we assemble
\be
\label{K matrix}
\sK=\begin{pmatrix}\sP&\so&\so\\\so&\sB&\sC\\\so&\sC\Tsf&\sB\end{pmatrix}.
\ee
In this new notation eq.~(\ref{altogether}) reads as an
$[3(L+1)^2-2]\times[3(L+1)^2-2]$-dimensional algebraic eigenvalue 
problem
\be\label{matrix eigenvalue problem}
 \sK\hsom\beg=\lambda\hsom\beg,
\ee
whose eigenvectors $\beg_1,\beg_2,\ldots,\beg_{3(L+1)^2-2}$ are mutually
orthogonal in the sense
\begin{equation}\label{spectral inner product}
\bega\Tsf\hsom\begb^{}=\dab, \qquad 
\bega\Tsf\sK\hsom\begb^{}=\lambda_\alpha\dab.
\end{equation}
The associated eigenfields
$\bg_1(\rhat),\bg_2(\rhat),\dots,\bg_{3(L+1)^2-2}(\rhat)$ are 
orthogonal over the region~$R$ and orthonormal over the whole
sphere~$\Omega$:  
\begin{equation}\label{spatial inner products}
\into \bga\cdot \bgb \domg=\dab,\qquad \intr \bga\cdot
\bgb \domg = \lambda_\alpha \dab. 
\end{equation}
The relations (\ref{spatial inner products}) for the spatial-domain
are equivalent to their matrix counterparts (\ref{spectral inner
  product}).  The eigenfield $\bg_1(\rhat)$ with the largest
eigenvalue~$\lambda_1$ is the element in the space of bandlimited
vector fields with most of its spatial energy within region~$R$; the
eigenfield $\bg_2(\rhat)$ is the next best-concentrated bandlimited
function orthogonal to~$\bg_1(\rhat)$ over both~$\Omega$ and~$R$; and
so on. Finally, as in the scalar case, we can sum up the eigenvalues
of the matrix~$\sK$ to define a vectorial spherical Shannon number
\ber\label{Shannon number}
 N\DDDD&=&\sumalpha \lambda_\alpha = \tr \sK= \sumshL
  (\Dlmlm+\Blmlm+\Clmlm) \\ 
  &=& \intr \left[\sumshL
  \bPlm(\rhat)\cdot \bPlm(\rhat)
  +\bBlm(\rhat)\cdot \bBlm(\rhat) 
+\bClm(\rhat)\cdot \bClm(\rhat)\right] \domg\\
  &=&\left[3(L+1)^2-2\right]\frac{A}{4\pi}.
\eer
To establish the last equality we used the
relation~(\ref{vecadd1}). Given the decoupling of the radial from the
tangential solutions that is apparent from eq.~(\ref{K matrix}) 
we may subdivide the vectorial spherical Shannon number into a radial
and a tangential one, $N^r=(L+1)^2A/(4\pi)$  and
$N^t=[2(L+1)^2-2]A/(4\pi)$,  respectively. 

\begin{figure}\center
\rotatebox{-90}{\includegraphics[height=0.8\columnwidth]
{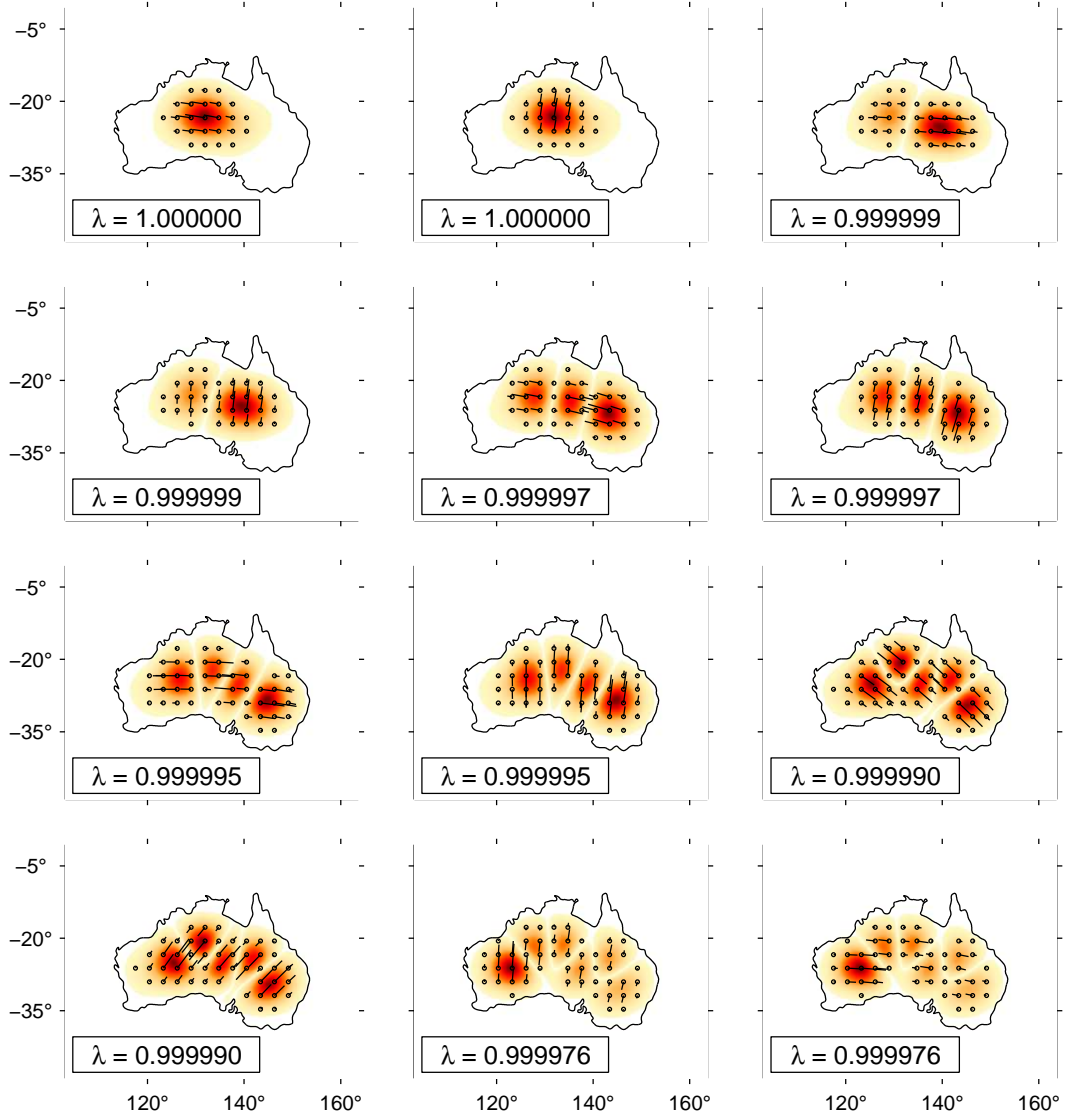}}
\caption{\label{australia} Twelve tangential Slepian functions
  $\bg_1,\bg_2,\ldots,\bg_{12}$, bandlimited to $L=60$, optimally
  concentrated within Australia. The concentration factors $\lambda_1,
  \lambda_2,\ldots,\lambda_{12}$ are indicated. The rounded tangential
  Shannon number $N^t=112$. Order of concentration is left to right,
  top to bottom. Color is absolute value (red the maximum) and circles
  with strokes indicate the direction of the eigenfield on the
  tangential plane. Regions in which the absolute value is less than
  one hundredth of the maximum absolute value on the sphere are left
  white.  }
\end{figure}

An example of tangential vectorial Slepian functions on a geographical
domain on the surface of the sphere is found in Figure~\ref{australia}.  

\sssec{Sturm-Liouville Character and Tridiagonal Matrix Formulation}

When the region of concentration $R$ is a symmetric polar cap with
colatitudinal radius~$\Theta$ centered on the north pole, special
rules apply that greatly facilitate the construction of the
localization kernel~(\ref{K matrix}). There are reductions of
eq.~(\ref{collectlater}) to some very manageable integrals that can be
carried out exactly by recursion. Solutions for the polar cap can be
rotated anywhere on the unit sphere using the same transformations
that apply in the rotation of scalar functions
~\cite[]{Blanco+97,Dahlen+98,Edmonds96,Freeden+2009}. 

In the axisymmetric case the matrix elements~(\ref{Dlmlmpagain}),
(\ref{Blmlmp  elements}) and~(\ref{Clmlmp elements}) reduce to   
\begin{eqnarray}\label{axi1}
 \Dlmlmp&=&2\pi\hsom\dmmp\intoth \Xlm\Xlpm \sindth,\\
% Check if the abs can be removed
 \Blmlmp&=&\frac{\displaystyle 2\pi\hsom\dmmp\intoth 
\left[\dXlm\dXlpm+m^2\divsinsq \Xlm \Xlpm \right] \sindth}{\displaystyle
   \renormprod},\label{axi2}\\ % Check the abs
\Clmlmp&=&-\frac{\displaystyle 2\pi\hsom\delta_{-mm'}\hsom m\intoth 
\left[\dXlm\Xlpm+ \Xlm \dXlpm \right] \dth}{\displaystyle
   \renormprod}
=
-\frac{\displaystyle 2\pi\hsom\delta_{-mm'}\hsom m
  \Xlm(\Theta)\Xlpm(\Theta)}{\displaystyle \renormprod} \label{axi3}
,
\end{eqnarray}
using the derivative notation $\dXlm= dX_{lm}/d\theta$ for the
normalized associated Legendre functions of
eq.~(\ref{xlm}). Eq.~(\ref{axi3}) can be easily evaluated. The
integrals over the product terms $\Xlm\Xlpm$ in eq.~(\ref{axi1}) can
be rewritten using Wigner $3j$
symbols~\cite[]{Wieczorek+2005,Simons+2006a,Plattner+2013} to simple integrals
over $X_{l\hsom 2m}$ or $X_{l\hsom 0}$ which can be handled
recursively~\cite[]{Paul78}. Finally, in eq.~(\ref{axi2}) integrals of
the type~$\dXlm \dXlpm$, and $m^2 \divsinsq \Xlm\Xlpm$ can be
rewritten as integrals over undifferentiated products of Legendre
functions~\cite[]{Ilk83,Eshagh2009a,Plattner+2013}. All in all, these
computations are straightforward to carry out and lead to
block-diagonal matrices at constant order~$m$, which are relatively
easily diagonalized. 

% So the separable form start writing +m and -m using cos and m. Carry
% it all through and take components together as Tony did. If you took
% the other one switch cos and sin with a minus and it's indeed a 90
% degree rotation of the ones that you had before. Seems trivial in
% retrospect. Alain also has a Preparations document to that
% effect. We don't much learn from this other than that the +m and -m
% are 90 degree rotations of each other. It is not like they neatly
% separate as in the radial case since they each have theta and phi
% components. WHAT IS THE NAME OF THAT DOCUMENT? I SEEM TO HAVE THROWN
% THIS OUT. IT ENDED WITH THE DESCRIPTION OF THE RADIAL AND TANGENTIAL
% COMPONENTS IN FUNCTION OF THE SIGN OF M AND HOW THEY SEPARATED. WAS
% QUITE NICE, ACTUALLY.   

As this chapter went to press, \cite{Jahn+2013b} reported the exciting
discovery of a differential operator that commutes with the tangential
part of the concentration operator~(\ref{nofurther}), and a
tridiagonal matrix formulation for the tangential vectorial
concentration problem to axisymmetric domains. They achieve this feat
by a change of basis by which to reduce the
vectorial problem to a scalar one that is separable in $\theta$ and
$\phi$,  using the special functions $\dXlm\pm m\hsom(\sin\theta)^{-1}\Xlm$
\cite[]{Sheppard+97}. Hence they derive a commuting differential
operator and a corresponding spectral matrix for the concentration
problem. By their approach, the solutions to the fixed-order
tangential concentration problem are again solutions to a
Sturm-Liouville problem with a very simple eigenvalue spectrum, and
the calculations are always fast and stable, much as they are for the
radial problem which completes the construction of vectorial Slepian
functions on the sphere. 

\ssec{Mid-Term Summary}

It is interesting to reflect, however heuristically, on the
commonality of all of the above aspects of spatiospectral
localization, in the slightly expanded context of reproducing-kernel
Hilbert
spaces~\cite[]{Nashed+91,Daubechies92,Yao67,Amirbekyan+2008b,Kennedy+2013}.
In one dimension, the Fourier orthonormality relation and the
``reproducing'' properties of the spatial delta function are given by
\be
\label{BFortho0}
\delta(t,t')=
\jnorm\intinf
e^{i\omega(t-t')}\,d\omega
,\qquad \intinf f(t')\hsp\delta(t,t')\,dt'=f(t)
.
\ee
In two Cartesian dimensions the equivalent relations are
\be
\label{BFortho00}
\delta(\bx,\bx')=\fnorm\intinf
e^{i\bk\cdot(\bx-\bx')}\dbk
,\qquad \intinf f(\bx')\hsp\delta(\bx,\bx')\dbx'=f(\bx)
,
\ee
and on the surface of the unit sphere we have, for the scalar case,
\be
\label{BFortho000}
\delta(\rhat,\rhat')=\sum_{l=0}^\infty\tlofp\!P_l(\rhat\cdot\rhat')
,\qquad
\into f(\rhat')\hsp\delta(\rhat,\rhat')\domg'=f(\rhat)
,
\ee
and for the vector case, we have the sum of dyads
\begin{subequations}
\label{BFortho0000}
\be
\label{Dirac delta expanded}
\bdelta(\rhat,\rhat')= 
\sumsh \bPlm(\rhat)\bPlm(\rhat')+
\bBlm(\rhat)\bBlm(\rhat') +
\bClm(\rhat)\bClm(\rhat'),
\ee
\be\label{Dirac delta}
 \into\bef(\rhat')\cdot\bdelta(\rhat,\rhat')\domg'=\bef(\rhat).
\ee
\end{subequations}
The integral-equation kernels~(\ref{slepian4}), (\ref{Beigen4}),
(\ref{banddelta}) and~(\ref{bandlimited Dirac}) are all bandlimited
spatial delta functions which are reproducing kernels for bandlimited
functions of the types in eqs~(\ref{gband}), (\ref{Bgdefn}),
(\ref{bandlg}) and~(\ref{bandlimited field}):   
\ber
\label{dttp}
\Dttp=\jnorm\intW 
e^{i\omega(t-t')}\,d\omega,\qquad&&\intinf g(t')\Dttp\,dt'=g(t),\\
\label{dxxp}
\Dxxp=
\fnorm
\intK e^{i\bk\cdot(\bx-\bx')}\dbk,\qquad&&\intinf
g(\bx')\Dxxp\dbx'=g(\bx),\\ 
\label{drrp}
\Drhrhp=\sum_{l=0}^L\tlofp\!P_l(\rhat\cdot\rhat'),\qquad&&\into
g(\rhat')\Drhrhp\domg=g(\rhat),\\[-1em]
\bmp[b]{0.35\textwidth}{
\ber
\lefteqn{\bDrhrhp=\sumshL\bPlm(\rhat)\bPlm(\rhat')}\hspace{2.5em}\nnr\\
&&+\bBlm(\rhat)\bBlm(\rhat')+\bClm(\rhat)\bClm(\rhat')\nnr
,
\eer
}\emp
\qquad&&
 \into\bg(\rhat')\cdot\bDrhrhp\domg'=\bg(\rhat).
\label{bigdrrp}
\eer
The equivalence of eq.~(\ref{dttp}) with eq.~(\ref{slepian4}) is
through the Euler identity and the reproducing properties follow from
the spectral forms of the orthogonality
relations~(\ref{BFortho0})--(\ref{BFortho00}), which are self-evident
by change-of-variables, and from the spectral form of
eq.~(\ref{BFortho000}), which is eq.~(\ref{normalization}). Much as
the delta functions of eqs~(\ref{BFortho0})--(\ref{BFortho0000}) set up
the Hilbert spaces of all square-integrable functions on the real
line, in two-dimensional Cartesian space, and on the surface of the
sphere (both scalar and vector functions), the
kernels~(\ref{dttp})--(\ref{bigdrrp}) induce the equivalent 
subspaces of bandlimited functions in their respective
dimensions. Inasmuch as the Slepian functions are the integral
eigenfunctions of these reproducing kernels in the sense of
eqs.~(\ref{eig1}), (\ref{eig2}), (\ref{eig3}), and~(\ref{nofurthera}),
they are complete bases for their band-\textit{limited} 
subspaces~\cite[]{Slepian+61,Landau+61,Daubechies92,Flandrin98,Freeden+98c,Plattner+2013}. 
Therein, the~$N\D$, $N\DD$, $N\DDD$ or $N\DDDD$ best time- or
space-\textit{concentrated} members allow for sparse, approximate,
expansions of signals that are spatially concentrated to
the one-dimensional interval $t\in[-T, T]\subset\mathbb{R}$, the
Cartesian region $\bx \in\mathcal{R}\subset \mathbb{R}^2$, or the
spherical surface patch $\rhat\in R\subset\Omega$.

As a corollary to this behavior, the infinite sets of \textit{exactly}
time- or spacelimited (and thus band-\textit{concentrated}) versions
of the functions~$g$, which are the eigenfunctions of
eqs~(\ref{dude1}), (\ref{Beigen34}), (\ref{firsttimeint})
and~(\ref{nofurther}) with the domains appropriately restricted, are
complete bases for square-integrable scalar functions on the intervals
to which they are
confined~\cite[]{Slepian+61,Landau+61,Simons+2006a,Plattner+2013}.
Expansions of such wideband signals in the small subset of their
$N\D$, $N\DD$, $N\DDD$ or $N\DDDD$ most band-concentrated members
provide reconstructions which are constructive in the sense that they
progressively capture all of the signal in the mean-squared sense, in
the limit of letting their numbers grow to infinity. This second class
of functions can be trivially obtained, up to a multiplicative
constant, from the bandlimited Slepian functions~$g$ by simple time-
or space limitation. While Slepian~\cite[]{Slepian+61,Slepian83}, for
this reason perhaps, never gave them a name, we have been referring to
those as~$h$ (and $\mbf{h}$) in our own investigations of similar
functions on the
sphere~\cite[]{Simons+2006a,Simons+2006b,Dahlen+2008,Plattner+2013}. 

% INTERESTING - SEE LANDAU67b, SUM OF EIGENVALUES SQUARED EQUALS
% DOUBLE INTEGRAL OF THE KERNEL, AS IN OUR COUPLING KERNEL. WHERE HAVE
% I SEEN THIS BEFORE IN THE THEORY OF SPECTRAL ANALYSIS? IN ONE OF THE
% PAPERS THAT LOOKED AT THE RATIO OF THE SUM OF THE EIGENVALUES TO THE
% SUM OF THE SQUARES OF THE EIGENVALUES. SOMEWHERE, NAMELY IN BRONEZ
% and LIU AND VAN VEEN, AND OF COURSE IMPLICIT IN OUR OWN WORK.  I
% ALSO SAW THIS RECENTLY AGAIN IN A LONG MATHEMATICAL PAPER BY I
% FORGET WHOM. SOMETHING CITED IN 2DCART IN THE YEARS 2000-2010 I
% BELIEVE, AND I ALSO BELIEVE I HAVE HIGHLIGHTED IT THERE.

% COMPRESSED SENSING ON THE SPHERE

% They are spatially doubly orthogonal, their Fourier expansions are
% orthogonal and they are inverse Fourier transforms of these orthogonal
% Fourier expansions, and thus nothing else than a recombination of
% the ``normal'' exponential basis functions that are the basis in L2.
% Mercer kernel. Separable Hilbert-Schmidt operator with an countably
% infinite set of eigenfunctions, that is what it is all about. The
% spectral theorem.

% Wolfgang Erb

%%%%%%%%%%%%%%%%%%%%%%%%%%%%%%%%%%%%%%%%%%%%%%%%%%%%%%%%%%%%%%%

\section{Problems in the Geosciences and Beyond}

Taking all of the above at face-value but referring again to the
literature cited thus far for proof and additional context, we return
to considerations closer to home, namely the estimation of geophysical
(or cosmological) signals and/or their power spectra, from noisy and
incomplete observations collected at or above the surface of the
spheres ``Earth'' or ``planet'' (or from inside the sphere ``sky'').
We restrict ourselves to real-valued scalar measurements, contaminated
by additive noise for which we shall adopt idealized models. We focus
exclusively on data acquired and solutions expressed on the
\textit{unit} sphere. We have considered generalizations to problems
involving satellite potential-field data collected at an altitude
elsewhere~\cite[]{Simons+2006b,Simons+2009b}. We furthermore note that
descriptions of the scalar gravitational and magnetic potential may be
sufficient to capture the behavior of the corresponding gravity and
magnetic vector fields, but that with vectorial Slepian functions
versatile and demanding satellite data analysis problems will be able
to get robustly handled even in the presence of noise that may be
strongly heterogeneous spatially and/or over the individual vector
components. Speaking quite generally, the two different statistical
problems that arise when geomathematical scalar spherical data are
being studied are, (i)~how to find the ``best'' estimate of the signal
given the data, and (ii)~how to construct from the data the ``best''
estimate of the power spectral density of the signal in question.
There are problems intermediate between either case, for instance
those that utilize the solutions to problems of the kind~(i) to make
inference about the power spectral density without properly solving
any problems of kind~(ii). Mostly such scenarios, e.g. in localized
geomagnetic field analysis~\cite[]{Beggan+2013} are born out of necessity
or convenience. We restrict our analysis to the ``pure'' end-member problems. 

Thus, let there be some real-valued scalar data distributed on the
unit sphere, consisting of ``signal'', $s$ and ``noise'', $n$, and let
there be some region of interest $R\subset\Omega$, in other words, let
\ber d(\rhat)=\left\{\begin{array}{ll}
    s(\rhat)+n(\rhat) & \mbox{if $\rhat\in R$},\\
    \mbox{unknown/undesired} & \mbox{if $\rhat\in \Omega\setminus R$}.
\end{array}\right.
\label{datdefs}
\eer
We assume that the signal of interest can be expressed by way of
spherical harmonic expansion as in eq.~(\ref{expansion}), and that it
is, itself, a realization of a zero-mean, Gaussian, isotropic, random
process, namely 
\be
s(\rhat)=\sumsh s_{lm}\Ylmrh, \qquad
\langle s_{lm}\rangle =0\also 
\langle s_{lm}s_{l'm'}\rangle=S_l\,\dllp\dmmp.
\ee
For illustration we furthermore assume that the noise is a zero-mean
stochastic process with an isotropic power spectrum, i.e. $\langle
n(\rhat)\rangle=0$ and $\langle
n_{lm}n_{l'm'}\rangle=N_l\,\dllp\dmmp$, and that it is 
statistically uncorrelated with the signal. We refer to 
power as \textit{white} when $S_l=S$ or $N_l=N$, or, equivalently,
when $\langle n(\rhat)n(\rhat')\rangle=N\delta(\rhat,\rhat')$. 
Our objectives are thus (i)~to determine the best
estimate~$\hat{s}_{lm}$ of the spherical harmonic expansion
coefficients~$s_{lm}$ of the signal and 
(ii)~to find the best estimate~$\hat{S}_l$ for the isotropic power
spectral density~$S_l$. While in the physical world there can be no limit
on bandwidth, practical restrictions force any and all of our
estimates to be bandlimited to some maximum spherical harmonic degree
$L$, thus by necessity $\hat{s}_{lm}=0$ and $\hat{S}_l=0$ for~$l> L$:
\be\label{wouldbe}
\hat{s}(\rhat)=\sumshL \hat{s}_{lm}\Ylmrh
.
\ee
This limitation, combined with the statements eq.~(\ref{datdefs}) on
the data coverage or the study region of interest, naturally leads us
back to the concept of ``spatiospectral concentration'', and, as we
shall see, solving either problem~(i) or~(ii) will gain from involving
the ``localized'' scalar Slepian functions rather than, or in addition
to, the ``global'' spherical harmonics basis.

This leaves us to clarify what we understand by ``best'' in this
context. While we adopt the traditional statistical metrics of bias,
variance, and mean-squared error to appraise the quality of our
solutions~\cite[]{Cox+74,Bendat+2000}, the resulting connections to
sparsity will be real and immediate, owing to the Slepian functions
being naturally instrumental in constructing
efficient, consistent and/or unbiased estimates of
either~$\hat{s}_{lm}$ or~$\hat{S}_l$. Thus, we define
\be
v=\langle\shat^2\rangle-\langle\shat\rangle^2
,\qquad 
b=\langle\shat\rangle-s,\qquad 
\epsilon=\shat-s,
\also
\langle\epsilon^2\rangle=v+b^2
\ee
for problem~(i), where the lack of subscript indicates that we can
study variance, bias and mean-squared error of the estimate of the
coefficients~$\hat{s}_{lm}$ but also of their spatial expansion
$\shat(\rhat)$. For problem~(ii) on the other hand, we focus 
on the estimate of the isotropic power spectrum at a given spherical
harmonic degree~$l$ by identifying 
\be
v_l=\langle\Shat^2_l\rangle-\langle\Shat_l\rangle^2,\qquad 
b_l=\langle\Shat_l\rangle-S_l,\qquad
\epsilon_l=\Shat_l-S_l,\also
\langle\epsilon^2_l\rangle=v_l+b^2_l
.
\ee
Depending on the application, the ``best'' estimate could mean the
unbiased one with the lowest
variance~\cite[]{Tegmark97b,Tegmark+97,Bond+98,Oh+99,Hinshaw+2003}, it
could be simply the minimum-variance estimate having some acceptable
and quantifiable bias~\cite[]{Wieczorek+2007}, or, as we would usually
prefer, it would be the one with the minimum mean-squared
error~\cite[]{Simons+2006b,Dahlen+2008}. 

\ssec{Problem~(i): Signal Estimation from Spherical Data}

\sssec{Spherical Harmonic Solution}

Paraphrasing results elaborated elsewhere~\cite[]{Simons+2006b}, we 
write the bandlimited solution to the damped inverse
problem 
\be
\intr(s-d)^2\domg+\eta\intbr s^2\domg=\mbox{minimum}
,
\label{variational}
\ee
where~$\eta\ge 0$ is a damping parameter, by straightforward algebraic
manipulation, as 
\be
\hat{s}_{lm}=\sumshLp \left(\Dlmlmp+\eta\bDlmlmp\right)^{-1} 
\intr d\,\Ylmp\domg
\label{hatss}
,
\ee
where~$\bDlmlmp$, the kernel that localizes to the
region~$\bar{R}=\Omega\setminus R$, compliments~$\Dlmlmp$ given by
eq.~(\ref{Dlmlmpdef}) which localizes to~$R$. Given the eigenvalue
spectrum of the latter, its inversion is inherently unstable, thus
eq.~(\ref{variational}) is an ill-conditioned inverse problem
unless~$\eta>0$, as has been well known, e.g. in
geodesy~\cite[]{Xu92a,Sneeuw+97}. Elsewhere~\cite[]{Simons+2006b} we have
derived exact expressions for the optimal value of the damping
parameter~$\eta$ as a function of the signal-to-noise ratio under
certain simplifying assumptions. As can be easily shown, without
damping the estimate is unbiased but effectively incomputable; the
introduction of the damping term stabilizes the solution at the cost
of added bias. And of course when~$R=\Omega$, eq.~(\ref{hatss}) is
simply \textit{the}  spherical harmonic transform, as in that case,
eq.~(\ref{Dlmlmpdef}) reduces to eq.~(\ref{normalization}), in other
words, then~$\Dlmlmp=\dllp\dmmp$.

\sssec{Slepian Basis Solution}

The trial solution in the Slepian basis designed for this
region of interest~$R$, i.e.
\be
\hat{s}(\rhat)=\sumapot\hat{s}_\alpha \garh 
,
\label{shatdef1}
\ee
would be completely equivalent to the expression in
eq.~(\ref{wouldbe}) by virtue of the completeness of the Slepian basis
for bandlimited functions everywhere on the sphere and the unitarity
of the transform~(\ref{bandlg}) from the spherical
harmonic to the Slepian basis. The solution to the undamped ($\eta=0$)
version of eq.~(\ref{variational}) would then be 
\be
\hat{s}_\alpha=\lambda_\alpha^{-1}\intr d\hsp g_\alpha\domg
,\label{SGsolspec}
\ee
which, being completely equivalent to eq.~(\ref{hatss}) for
$\eta=0$, would be computable, and biased, only when the expansion in
eq.~(\ref{shatdef1}) were to be truncated to some finite
$J<\Lpot$ to prevent the blowup of the eigenvalues~$\lambda$. Assuming
for simplicity of the argument that~$J=N\DDD$, the 
essence of the approach is now that the solution 
\be
\hat{s}(\rhat)=\sumaN\hat{s}_\alpha \garh
\label{shatdef2}
\ee
will be sparse (in achieving a bandwidth~$L$ using~$N\DDD$
Slepian instead of~$\Lpot$ spherical-harmonic expansion coefficients)
yet good (in approximating the signal as well as possible in the
mean-squared sense in the region of interest~$R$) and of geophysical
utility (assuming we are dealing with spatially localized processes
that are to be extracted, e.g., from global satellite
measurements) as shown by~\cite{Han+2008a}, \cite{Simons+2009b},
and~\cite{Harig+2012}. 

\ssec{Bias and Variance}

In concluding this section let us illustrate another welcome
by-product of our methodology, by writing the mean-squared error for
the spherical harmonic solution~(\ref{hatss}) compared to the
equivalent expression for the solution in the Slepian basis,
eq.~(\ref{SGsolspec}). We do this as a function of the spatial
coordinate, in the Slepian basis for both, and, for maximum clarity of
the exposition, using the contrived case when both signal and noise
should be white (with power $S$ and $N$, respectively) as well as
bandlimited (which technically is impossible). In the former case, we 
get
\ber
\label{msefinal}
\langle\epsilon^2(\rhat)\rangle&=&N\sumapot\lambda_\alpha
[\lambda_\alpha+\eta(1-\lambda_\alpha)]^{-2}
g_\alpha^2(\rhat)\\
&&{}+\eta^2 S\sumapot
(1-\lambda_\alpha)^2[\lambda_\alpha+\eta(1-\lambda_\alpha)]^{-2}
g_\alpha^2(\rhat)\nnr 
,
\eer
while in the latter, we obtain
\be
\langle\epsilon^2(\rhat)\rangle=N\sumaN\lambda_\alpha^{-1}g_\alpha^2(\rhat)+S\sumakR
g_\alpha^2(\rhat)
.
\label{SGmsefinal}
\ee
All~$\Lpot$ basis functions are required to express the mean-squared
estimation error, whether in eq.~(\ref{msefinal}) or in
eq.~(\ref{SGmsefinal}). The first term in both expressions is the
variance, which depends on the measurement noise. Without damping or truncation
the variance grows without bounds. Damping and truncation alleviate
this at the expense of added bias, which depends on the
characteristics of the signal, as given by the second term. In contrast
to eq.~(\ref{msefinal}), however, the Slepian
expression~(\ref{SGmsefinal}) has disentangled the contributions due
to noise/variance and signal/bias by projecting them onto the sparse
set of well-localized and the remaining set of poorly localized Slepian
functions, respectively. The estimation variance is felt via the
basis functions $\alpha=1\rar N\DDD$ that are well concentrated inside
the measurement area, and the effect of the bias is relegated to those
$\alpha=N\DDD+1\rar\Lpot$ functions that are confined to the region of
missing data. 

When forming a solution to problem~(i) in the Slepian basis by
truncation according to~eq.~(\ref{shatdef2}), changing the truncation
level~$J$ to values lower or higher than the Shannon number~$N\DDD$
amounts to navigating the trade-off space between variance, bias (or
``resolution''), and sparsity in a manner that is captured with great
clarity by eq.~(\ref{SGmsefinal}). We refer the reader
elsewhere~\cite[]{Simons+2006b,Simons+2007} for more details, and, in particular, for
the case of potential fields estimated from data collected at
satellite altitude.

\ssec{Problem~(ii): Power Spectrum Estimation from Spherical Data}

Following~\cite{Dahlen+2008} we will find it convenient to
regard the data~$d(\rhat)$ given in eq.~(\ref{datdefs}) as having been
multiplied by a unit-valued boxcar window function confined to the
region~$R$,   
\be
\label{boxcar}
b(\rhat)=\sum_{p=0}^\infty\sum_{q=-p}^pb_{pq}Y_{pq}(\rhat)
=\left\{\begin{array}{ll} 
1 & \mbox{if~$\rhat\in R$},\\
0 & \mbox{otherwise.} \end{array}\right.
\ee
The power spectrum of the boxcar window~(\ref{boxcar}) is 
\be \label{boxspec}
B_p=\frac{1}{2p+1}\sum_{q=-p}^pb_{pq}^2.
\ee

\sssec{The Spherical Periodogram}

Should we decide that an acceptable estimate of the power spectral
density of the available data is nothing else but the weighted average of
its spherical harmonic expansion coefficients, we would be forming the
spherical analogue of what Schuster~\cite[]{Schuster1898} named the
``periodogram'' in the context of time series analysis, namely
\be \label{SestSP} \hat{S}_l\SP=
\left(\frac{4\pi}{A}\right)\frac{1}{2l+1}\sum_{m=-l}^l
\left[\int_Rd(\rhat)\,\Ylmrh\domg
\right]^2
.
\ee

\sssec{Bias of the Periodogram}

Upon doing so we would discover that the expected value of
such an estimator would be the biased quantity
\be
\label{SexpecSP}
\langle \hat{S}_l\SP\rangle
=\sum_{l'=0}^\infty K_{ll'}(S_{l'}+N_{l'}),
\ee
where, as it is known in astrophysics and
cosmology~\cite[]{Peebles73,Hauser+73,Hivon+2002}, the
periodogram ``coupling'' matrix
\be \label{Kmatdef}
K_{ll'}=\left(\frac{4\pi}{A}\right)\frac{1}{2l+1}
\sum_{m=-l}^l\sum_{m'=-l'}^{l'} 
\left[D_{lm,l'm'}\right]^2
,
\ee
governs the extent to which an estimate $\hat{S}_l\SP$ of~$S_l$ is
influenced by  spectral leakage from  power in
neighboring spherical harmonic degrees $l'=l\pm 1,l\pm 2,\ldots$, all
the way down to 0 and up to~$\infty$. In
the case of full data coverage, $R=\Omega$, or of a perfectly white
spectrum, $S_l=S$, however, the estimate would be unbiased ---
provided the noise spectrum, if known, can be subtracted beforehand. 

\sssec{Variance of the Periodogram}

The covariance of the periodogram estimator~(\ref{SestSP}) would moreover be
suffering from strong wideband coupling of the power spectral
densities in being given by  
\be \label{SigmaSP2}
\Sigma_{ll'}\SP=\frac{2(4\pi/A)^2}{(2l+1)(2l'+1)}\sum_{m=-l}^l\sum_{m'=-l'}^{l'}
\left[\sum_{p=0}^{\infty}\sum_{q=0}^{\infty}(S_p+N_p)
D_{lm,pq}D_{pq,l'm'}\right]^2.
\ee
Even under the commonly made assumption as should the power spectrum
be slowly varying within the main lobe of the coupling matrix, such
coupling would be nefarious. In the ``locally white'' case we would have
\be \label{moderSP}
\Sigma_{ll'}\SP=\frac{2(4\pi/A)^2}{(2l+1)(2l'+1)}
\,(S_l+N_l)(S_{l'}+N_{l'})\sum_{m=-l}^l\sum_{m'=-l'}^{l'}
\left[D_{lm,l'm'}\right]^2.
\ee
Only in the limit of whole-sphere data coverage will
eqs~(\ref{SigmaSP2}) or~(\ref{moderSP}) reduce to  
\be \label{SigmaWSfin}
\Sigma_{ll'}\WS=\frac{2}{2l+1}\left(S_l+N_l\right)^2\dllp,
\ee
which is the ``planetary'' or  ``cosmic'' variance that can be
understood on the basis of elementary statistical
considerations~\cite[]{Jones63,Knox95,Grishchuk+97}. The strong
spectral leakage for small regions ($A\ll 4\pi$) is highly
undesirable and makes the periodogram `hopelessly
obsolete'~\cite[]{Thomson+91}, or, to put it kindly,
`naive'~\cite[]{Percival+93}, just as it is for one-dimensional time
series. 

In principle it is possible  --- after subtraction of the noise
bias --- to eliminate the leakage bias in the
periodogram estimate~(\ref{SestSP}) by numerical inversion of the
coupling matrix~$K_{ll'}$. Such a 
``deconvolved periodogram'' estimator is unbiased. 
However, its covariance depends on the inverse of the periodogram
coupling matrix, which is only feasible when the region~$R$
covers most of the sphere, $A\approx 4\pi$. For any region whose area
$A$ is significantly smaller than~$4\pi$, the periodogram coupling
matrix~(\ref{Kmatdef}) will be too ill-conditioned to be invertible.

Thus, much like in problem~(i) we are faced with bad bias and poor
variance, both of which are controlled by the lack of localization of
the spherical harmonics and their non-orthogonality over incomplete
subdomains of the unit sphere. Both effects are described by the
spatiospectral localization kernel defined in~(\ref{Dlmlmpdef}),
which, in the quadratic estimation problem~(ii) appears in ``squared''
form in eq.~(\ref{SigmaSP2}). Undoing the effects of the wideband
coupling between degrees at which we seek to estimate the power
spectral density by inversion of the coupling kernel is virtually
impossible, and even if we could accomplish this to remove the
estimation bias, this would much inflate the estimation
variance~\cite[]{Dahlen+2008}. 

\sssec{The Spherical Multitaper Estimate}

We therefore take a page out of the one-dimensional power estimation
playbook of~\cite{Thomson82} by forming the ``eigenvalue-weighted
multitaper estimate''.  We could weight single-taper estimates
adaptively to minimize quality measures such as estimation variance or
mean-squared error ~\cite[]{Thomson82,Wieczorek+2007}, but in
practice, these methods tend to be rather computationally
demanding. Instead we simply multiply the data~$d(\rhat)$ by the
Slepian functions or ``tapers''~$g_{\alpha}(\rhat)$ designed for the
region of interest prior to computing power and then averaging:   
\be \label{SMTdef}
\hat{S}_l\MT=\sum_{\alpha=1}^{\Lpot}\lambda_\alpha
\left(\frac{4\pi}{N\DDD}\right)\frac{1}{2l+1}
\sum_{m=-l}^l\left[\int_{\Omega} g_{\alpha}(\rhat)\,
d(\rhat)\,Y_{lm}(\rhat)\domg\right]^2.
\ee

\sssec{Bias of the Multitaper Estimate}

The expected value of the estimate~(\ref{SMTdef}) is
\be \label{SMTexpec}
\langle\hat{S}_l\MT\rangle=\sum_{l'=l-L}^{l+L}M_{ll'}(S_{l'}+N_{l'})
,
\ee
where the eigenvalue-weighted multitaper coupling matrix,
using Wigner 3-$j$ functions~\cite[]{Varshalovich+88,Messiah2000}, is
given by
\be \label{MTcouple}
M_{ll'}=\frac{2l'+1}{\Lpot}\sum_{p=0}^L(2p+1)
\!\left(\!\begin{array}{ccc}
l & p & l' \\ 0 & 0 & 0\end{array}\!\right)^2.
\ee
It is remarkable that this result depends only upon the chosen bandwidth~$L$
and is completely independent of the size, shape or connectivity of
the region~$R$, even as~$R=\Omega$. Moreover, every row of the matrix
in eq.~(\ref{MTcouple}) sums to unity, which ensures that a
the multitaper spectral estimate $\hat{S}_l\MT$ has no leakage
bias in the case of a perfectly white spectrum provided the noise bias
is subtracted as well: $\langle\hat{S}_l\MT\rangle-\sum
M_{ll'}N_{l'}=S$ if~$S_l=S$.  

\sssec{Variance of the Multitaper Estimate}

Under the moderately colored approximation, which is more easily
justified in this case because the coupling~(\ref{MTcouple}) is
confined to a narrow band of width 
less than or equal to~$2L+1$, with~$L$ the bandwidth of the tapers,
the eigenvalue-weighted multitaper covariance is 
\be \label{thisisit}
\Sigma_{ll'}\MT=\frac{1}{2\pi}(S_l+N_l)(S_{l'}+N_{l'})\sum_{p=0}^{2L}(2p+1)
\,\Gamma_p\!\left(\!\begin{array}{ccc}
l & p & l' \\ 0 & 0 & 0\end{array}\!\right)^2,
\ee
where, using Wigner 3-$j$ and 6-$j$
functions~\cite[]{Varshalovich+88,Messiah2000}, 
\ber \label{Gammafin}
\Gamma_p&=&\frac{1}{(N\DDD)^2}\sum_{s=0}^L\sum_{s'=0}^L
\sum_{u=0}^L\sum_{u'=0}^L 
(2s+1)(2s'+1)(2u+1)(2u'+1)\nnr\\
&&{}\times\sum_{e=0}^{2L}(-1)^{p+e}(2e+1)B_e \hspace{8em}\nonumber\\
&&{}\times\left\{\!\begin{array}{ccc}
s & e & s' \\ u & p & u'\end{array}\!\right\}\left(\!\begin{array}{ccc}
s & e & s' \\ 0 & 0 & 0\end{array}\!\right)\!\left(\!\begin{array}{ccc}
u & e & u' \\ 0 & 0 & 0\end{array}\!\right)\!\left(\!\begin{array}{ccc}
s & p & u' \\ 0 & 0 & 0\end{array}\!\right)\!\left(\!\begin{array}{ccc}
u & p & s' \\ 0 & 0 & 0\end{array}\!\right).
\eer
In this expression $B_e$, the boxcar power~(\ref{boxspec}),  which we
note \textit{does} depend on the shape of the region of interest, appears again,
summed over angular degrees limited by 3-$j$ selection rules to $0\leq
e\leq 2L$. The sum in eq.~(\ref{thisisit}) is likewise limited to
degrees $0\leq p\leq 2L$. The effect of tapering with windows
bandlimited to~$L$ is to introduce covariance between the estimates at
any two different degrees~$l$ and~$l'$ that are separated by fewer
than $2L+1$ degrees. Eqs~(\ref{thisisit})--(\ref{Gammafin}) are very
efficiently computable, which should make them competitive with, e.g.,
jackknifed estimates of the estimation variance
~\cite[]{Chave+87,Thomson+91,Thomson2007}.  

The crux of the analysis lies in the fact that
the  matrix of the spectral covariances between
\textit{single-}tapered estimates is almost
diagonal~\cite[]{Wieczorek+2007}, showing   
that the individual estimates that enter the weighted formula~(\ref{SMTdef})
are almost uncorrelated statistically. This embodies the very essence
of the multitaper method. It dramatically reduces the estimation
variance at the cost of small increases of readily quantifiable bias.

\section{Practical Considerations}

%\ssec{Notation and preliminary considerations}

In this section we now turn to the very practical context of sampled,
e.g. geodetic, data on the sphere. We shall deal exclusively with
bandlimited scalar functions, which are equally well expressed in the
spherical harmonic as the Slepian basis, namely:  
\be\label{bandlf1} 
f(\rhat)=\sumshL f_{lm}\Ylmrh=\sumapot f_\alpha\hsp \garh
,
\ee
whereby the Slepian-basis expansion coefficients are obtained as
\be\label{bandlf2}
f_\alpha=\into f(\rhat)\garh\domg
.
\ee
If the function of interest is spatially localized in the region~$R$,
a truncated reconstruction using Slepian functions built for the
same region will constitute a very good, and sparse, local
approximation to it~\cite[]{Simons+2009b}:  
\be\label{bandlfcon}
f(\rhat)\approx\sumaN f_\alpha\hsp \garh,\qquad\rhat\in R
.
\ee
We represent any sampled, bandlimited function $f$ by an
$M$-dimensional column vector  
\be\label{beff}
\beff=(f_1\; \cdots \; f_j \; \cdots\; f_M)\Tit
,
\ee
where $f_j=f(\br_j)$ is the value of $f$ at pixel $j$,
and $M$ is the total number of sampling locations. In the most 
general case the distribution of pixel centers will be completely
arbitrary~\cite[]{Hesse+2010}. The special case of equal-area
pixelization of a 2-D function $f(\br)$ on the unit sphere $\Omega$ is
analogous to the equispaced digitization of a 1-D time
series. Integrals will then be assumed to be approximated with
sufficient accuracy by a Riemann sum over a dense set of pixels,
\be \label{pixelsum}
\int\! f(\br)\domg\approx \Delta\Omega\sum_{j=1}^M f_j
\also
\int\! f^2(\br)\domg\approx\Delta\Omega\,\beff\T\beff
.
\ee
We have deliberately left the integration domain out of the above
equations to cover both the cases of sampling over the entire unit
sphere surface~$\Omega$, in which case the solid angle
$\Delta\Omega=4\pi/M$ (case~1) as well as over an incomplete
subdomain~$R\subset\Omega$, in which case $\Delta\Omega=A/M$, with $A$
the area of the region~$R$ (case~2).  If we collect  the real spherical
harmonic basis functions $\Ylm$ into an $\Lpot\times M$-dimensional matrix  
\be
\bY=
\left(\barray{ccccc}
Y_{00}(\rhat_1) & \cdots     & Y_{00}(\rhat_j)  & \cdots &  Y_{00}(\rhat_M)\\
                &            & \vdots           &        &       \\
\cdots          &            & \Ylm(\rhat_j)    &        &\cdots \\
                &            & \vdots           &        &       \\
Y_{LL}(\rhat_1) & \cdots     & Y_{LL}(\rhat_j)  & \cdots &Y_{LL}(\rhat_M)       
\earray
\right)
,
\ee
and the spherical harmonic coefficients of the function into an
$\Lpot\times 1$-dimensional vector 
\be
\label{fdef}
\fb=(\;f_{00}\;\cdots\; f_{lm}\; \cdots\; f_{LL}\;)\Tit
,
\ee
we can write the spherical harmonic synthesis in eq.~(\ref{bandlf1})
for sampled data without loss of generality as 
\be
\label{synth}
\beff=\bY\T \fb
.
\ee
We will adhere to the notation convention of using sans-serif fonts
(e.g. $\beff$, $\bY$) for vectors or matrices that depend on at least
one spatial variable, and serifed fonts (e.g. $\fb,\Db$\hsp) for those
that are entirely composed of ``spectral'' quantities. In the case of
dense, equal-area, whole-sphere sampling we have an 
approximation to eq.~(\ref{normalization}):
\be
\label{nohold}
\bY\bY\T\approx\Delta\Omega^{-1}\Ib\qquad \mbox{(case~1)}
,
\ee
where the elements of the $\Lpot\times\Lpot$-dimensional spectral
identity matrix~$\Ib$ are given by the Kronecker deltas
$\dllp\dmmp$. In the case of dense, equal-area, sampling over some closed
region~$R$, we find instead an approximation to the  $\Lpot\times
\Lpot$-dimensional ``spatiospectral localization matrix'':
\be
\label{nohold2}
\bY\bY\T\approx\Delta\Omega^{-1}\Db\qquad \mbox{(case 2)}
,
\ee
where the elements of~$\Db$ are those defined in
eq.~(\ref{Dlmlmpdef}). 

Let us now introduce the $\Lpot\times \Lpot$-dimensional matrix of
spectral Slepian eigenfunctions by 
\be
\Gb=\left(\barray{ccccc}
g_{00\hsp1}       & \cdots     & g_{00\hsp\alpha}  & \cdots &  g_{00\hsp\Lpot}\\
                 &            & \vdots           &        &       \\
\cdots           &            & g_{lm\hsp\alpha}    &        &\cdots \\
                 &            & \vdots           &        &       \\
g_{LL\hsp 1} & \cdots     & g_{LL\hsp\alpha}  & \cdots &g_{LL\hsp\Lpot}
\earray
\right)
.
\ee
This is the matrix that contains the eigenfunctions of the problem
defined in eq.~(\ref{fulleigen1}), which we rewrite as
\be
\Db\hsp\Gb=\Gb\hsp\blambda
,
\ee
where the diagonal matrix with the concentration eigenvalues is given by
\be
\blambda=\diag\left(\;\lambda_1\;\cdots\;\lambda_\alpha\;\cdots\;\lambda_{\Lpot}\;\right)
.
\ee
The spectral orthogonality relations of eq.~(\ref{orthogn}) are
\be
\label{sleportho}
\Gb\Trm\Gb=\Ib,\qquad
\Gb\Trm\Db\hsp\Gb=\blambda
,
\ee
where the elements of the $\Lpot\times\Lpot$-dimensional Slepian
identity matrix~$\Ib$ are given by the Kronecker deltas
$\dab$. We write the Slepian functions of eq.~(\ref{bandlg}) as
\be
\label{slepwrite}
\begg=\Gb\Trm\bY
\also
\bY=\Gb\hsp\begg
,
\ee
where the $\Lpot\times M$-dimensional matrix holding the sampled
spatial Slepian functions is given by
\be
\begg=\left(\barray{ccccc}
g_{1}(\rhat_1) & \cdots     & g_{1}(\rhat_j)  & \cdots &  g_{1}(\rhat_M)\\
                &            & \vdots           &        &       \\
\cdots          &            & g_\alpha(\rhat_j)    &        &\cdots \\
                &            & \vdots           &        &       \\
g_{\Lpot}(\rhat_1) & \cdots     & g_{\Lpot}(\rhat_j)  & \cdots &g_{\Lpot}(\rhat_M)       
\earray
\right)
.
\ee
Under a dense, equal-area, whole-sphere sampling, we will
recover the the spatial orthogonality of eq.~(\ref{orthog})
approximately as
\be
\label{ggt1}
\begg\hsp\begg\T\approx\Delta\Omega^{-1}\Ib\qquad \mbox{(case~1)}
,
\ee
whereas for dense, equal-area, sampling over a region~$R$ we will get,
instead, 
\be
\label{ggt2}
\begg\hsp\begg\T\approx\Delta\Omega^{-1}\blambda\qquad \mbox{(case 2)}
.
\ee
With this matrix notation we shall revisit both estimation problems of
the previous section.

\ssec{Problem~(i), Revisited}

\sssec{Spherical Harmonic Solution}

If we treat eq.~(\ref{synth}) as a noiseless inverse
problem in which the sampled data~$\beff$ are given but from which
the coefficients~$\fb$ are to be determined, we find that for dense,
equal-area, whole-sphere sampling, the solution 
\be
\label{inv0}
\fbh\approx\Delta\Omega\,\bY\hsp\beff
\qquad\mbox{(case~1)}
\ee
is simply the discrete approximation to the spherical harmonic
analysis formula~(\ref{expansion}). For dense, equal-area, regional
sampling we need to calculate
\be
\label{inv2}
\fbh\approx\Delta\Omega\,\Db^{-1}\hsp\bY\beff
\qquad\mbox{(case 2)}
.
\ee
Both of these cases are simply the relevant solutions to the familiar
overdetermined spherical harmonic inversion
problem~\cite[]{Kaula67a,Menke89,Aster+2005a} for discretely sampled
data, i.e. the least-squares solution to eq.~(\ref{synth}),
\be
\label{inv1}
\fbh=(\bY\bY\T)^{-1}\bY\hsp\beff
,
\ee
for the particular cases described by
eqs~(\ref{nohold})--(\ref{nohold2}). In eq.~(\ref{inv2}) we
furthermore recognize the discrete version of eq.~(\ref{hatss}) with
$\eta=0$, the undamped solution to the minimum mean-squared error
inverse problem posed in continuous form in
eq.~(\ref{variational}). From the continuous limiting 
case eq.~(\ref{hatss}) we thus discover the general form that
damping should take in regularizing the ill-conditioned inverse required in
eqs~(\ref{inv2})--(\ref{inv1}). Its principal property is that it
differs from the customary \textit{ad hoc} practice of adding small
values on the diagonal only.  Finally, in the most general and
admittedly most commonly encountered case of randomly scattered data
we require the Moore-Penrose pseudo-inverse   
\be
\label{invgen}
\fbh=\pinv(\bY\T)\hsp\beff
,
\ee
which is constructed by inverting the singular value decomposition
(svd) of~$\bY\T$ with its singular values truncated beyond where they
fall below a certain threshold~\cite[]{Xu98}. Solving eq.~(\ref{invgen})
by truncated svd is equivalent to inverting a truncated eigenvalue
expansion of the normal matrix~$\bY\bY\T$ as it appears in
eq.~(\ref{inv1}), as can be easily shown.
% FJS
% But wouldn't we need to
% put in a weight matrix to take care
% of the metric, i.e. the sampling density??? I would think so. Or is
% this automatically taken care of since near duplicate sampling
% points will lead to linear dependence and thus be obliterated from
% the truncated solution? 

\sssec{Slepian Basis Solution}

If we collect the Slepian expansion coefficients of the
function~$f$ into the $\Lpot\times 1$-dimensional vector
\be
\label{tdef}
\tb=(\;f_{1}\;\cdots\; f_{\alpha}\; \cdots\; f_{\Lpot}\;)\Tit
,
\ee
the expansion~(\ref{bandlf1}) in the Slepian basis takes the form
\be
\label{synthslep}
\beff=\begg\T\tb=\bY\T\Gb\hsp\tb
,
\ee
where we used eqs~(\ref{sleportho})--(\ref{slepwrite}) to obtain the
second equality. Comparing eq.~(\ref{synthslep}) with
eq.~(\ref{synth}), we see that the Slepian expansion coefficients of a
function transform to and from the spherical harmonic coefficients as: 
\be\label{sleptoharm}
\fb=\Gb\hsp\tb
\also
\tb=\Gb\Trm\fb
.
\ee 
Under dense, equal-area, sampling with complete coverage
the coefficients in eq.~(\ref{synthslep}) can be estimated from
\be
\label{invg}
\tbh\approx\Delta\Omega\,\begg\hsp\beff
\qquad\mbox{(case~1)}
,
\ee
the  discrete, approximate, version of eq.~(\ref{bandlf2}). 
For dense, equal-area, sampling in a limited region~$R$ we get
\be
\label{invg2}
\tbh\approx\Delta\Omega\,
\blambda^{-1}\hsp\begg\hsp\beff
\qquad\mbox{(case 2)}
.
\ee
As expected, both of the solutions~(\ref{invg})--(\ref{invg2}) are again
special cases of the overdetermined least-squares solution
\be
\label{invg4}
\tbh=
(\begg\begg\T)^{-1}
\begg\hsp\beff
,
\ee
as applied to eqs~(\ref{ggt1})--(\ref{ggt2}).  We encountered
eq.~(\ref{invg2}) before in the continuous form of
eq.~(\ref{SGsolspec}); it solves the undamped minimum mean-squared
error problem~(\ref{variational}).  The regularization of
this ill-conditioned inverse problem may be achieved by truncation of the
concentration eigenvalues, e.g. by restricting the size of the
$\Lpot\times\Lpot$-dimensional operator~$\begg\begg\T$ to its first
$J\times J$ subblock. Finally,  in the most general, 
scattered-data case, we would be using an eigenvalue-truncated version
of eq.~(\ref{invg4}), or, which is equivalent, form the
pseudo-inverse  
\be
\label{invg3}
\tbh=\pinv(\begg\T)\hsp\beff
.
\ee
%%\textbf{INVESTIGATE HERE WHERE IT IS KNOWN THAT THE EIGENVALUES SHOULD
%%  BE STOPPED AT N/S AS WE ALREADY KNOW. COMMENT ON THE TRUNCATION
%%  ABOVE. See Jordan and Minster 1972, etc. And P. Xu.}
%% NOPE - THIS HERE TOO, IS EXACTLY EQUIVALENT TO THE TRUNCATED
%% EIGENVALUE EXPANSION - PINV OF G WOULD JUST YIELD LAMBDA AS THE
%% EIGENVECTORS ARE THE IDENTITY MATRIX - THE WHOLE THING IS ALREADY
%% DIAGONAL. OR ALMOST.

The solutions~(\ref{inv0})--(\ref{inv1})
and~(\ref{invg})--(\ref{invg4}) are equivalent and differ only by the
orthonormal change of basis from the spherical harmonics to the
Slepian functions. Indeed, using eqs~(\ref{slepwrite})
and~(\ref{sleptoharm}) to transform eq.~(\ref{invg4}) into an equation
for the spherical harmonic coefficients, and comparing with
eq.~(\ref{inv1}) exposes the relation 
\be\label{trivial}
\Gb(\begg\begg\T)^{-1}\Gb\Trm=(\bY\bY\T)^{-1}
,
\ee
which is a trivial identity for case~1 (insert eqs~\ref{nohold},
\ref{ggt1} and~\ref{sleportho}) and,  after substituting
eqs~(\ref{nohold2}) and~(\ref{ggt2}), entails
\be
\Gb \blambda^{-1}\Gb\Trm=\Db^{-1}
\ee
for case~2, which holds by virtue of
eq.~(\ref{sleportho}). Eq.~(\ref{trivial}) can also be
verified directly from eq.~(\ref{slepwrite}), which implies
\be
\bY\bY\T=
\Gb(\begg\begg\T)\hsp\Gb\Trm
.
\ee
The popular but labor-intensive procedure by which the unknown
spherical harmonic expansion coefficients of a scattered data set are
obtained by forming the Moore-Penrose pseudo-inverse as in
eq.~(\ref{invgen}) is thus equivalent to determining the truncated
Slepian solution of eq.~(\ref{invg2}) in the limit of continuous and
equal-area, but incomplete data coverage. In that limit, the generic
eigenvalue decomposition of the normal matrix becomes a specific
statement of the Slepian problem as we encountered it before, namely
\be
\bY\bY\T\hspm\Delta\Omega=\Ub\bsigma^2\Ub\Trm
\quad\rar\quad
\Db=\Gb\blambda\Gb\Trm
.
\ee
Such a connection has been previously pointed out for time
series~\cite[]{Wingham92} and leads to the notion of ``generalized
prolate spheroidal functions''~\cite[]{Bronez88} should the ``Slepian''
functions be computed from a formulation of the concentration problem
in the scattered data space directly, rather than being determined by
sampling those  obtained from solving the corresponding continuous
problem, as we have described here.  % SHOULD LOOK THIS UP 

Above, we showed how to stabilize the inverse problem of
eq.~(\ref{inv1}) by damping. We dealt with the case of continuously
available data only; the form in which it appears in eq.~(\ref{hatss})
makes it clear that damping is hardly practical for scattered data.
Indeed, it requires knowledge of the complementary localization
operator~$\bar{\Db}$, in addition to being sensitive to the choice
of~$\eta$, whose optimal value depends implicitly on the unknown
signal-to-noise ratio~\cite[]{Simons+2006b}. The data-driven approach
taken in eq.~(\ref{invgen}) is the more sensible one~\cite[]{Xu98}. We
have now seen that, in the limit of continuous partial coverage, this
corresponds to the optimal solution of the problem formulated directly
in the Slepian basis. It is consequently advantageous to also work in
the Slepian basis in case the data collected are scattered but closely
collocated in some region of interest. Prior knowledge of the geometry
of this region and a prior idea of the spherical harmonic bandwidth of
the data to be inverted allows us to construct a Slepian basis for the
situation at hand, and the problem of finding the Slepian expansion
coefficients of the unknown underlying function can be solved using
eqs~(\ref{invg4})--(\ref{invg3}). The measure within which this
approach agrees with the theoretical form of eq.~(\ref{invg2}) will
depend on how regularly the data are distributed within the region of
study, i.e on the error in the approximation~(\ref{ggt2}). But if
indeed the scattered-data Slepian normal matrix~$\begg\begg\T$ is
nearly diagonal in its first $J\times J$-dimensional block due to the
collocated observations having been favorably, if irregularly,
distributed, then eq.~(\ref{invg2}), which, strictly speaking,
requires no matrix inversion, can be applied directly. If this is not
the case, but the data are still collocated or we are only interested
in a local approximation to the unknown signal, we can
restrict~$\begg$ to its first~$J$ rows, prior to
diagonalizing~$\begg\begg\T$ or performing the svd of a
partial~$\begg\T$ as necessary to calculate
eqs~(\ref{invg4})--(\ref{invg3}).  Compared to solving
eqs~(\ref{inv1})--(\ref{invgen}), the computational savings will still
be substantial, as only when~$R\approx\Omega$ will the
operator~$\bY\bY\T$ be nearly diagonal. Truncation of the eigenvalues
of~$\bY\bY\T$ is akin to truncating the matrix~$\begg\begg\T$ itself,
which is diagonal or will be nearly so. With the theoretically,
continuously determined, sampled Slepian functions as a
parametrization, the truncated expansion is easy to obtain and the
solution will be locally faithful within the region of interest~$R$.
In contrast, should we truncate~$\bY\bY\T$ itself, without first
diagonalizing it, we would be estimating a low-degree approximation of
the signal which would have poor resolution
everywhere. See~\cite{Slobbe+2012} for a set of examples in a slightly
expanded and numerically more challenging context. 

\sssec{Bias and Variance}

For completeness we briefly return to the expressions for the
mean-squared estimation error of the damped spherical-harmonic and
the truncated Slepian function methods,
eqs.~(\ref{msefinal})--(\ref{SGmsefinal}), which we quoted for the
example of ``white'' signal and noise with power $S$ and $N$,
respectively. Introducing the   
$\Lpot\times \Lpot$-dimensional spectral matrices  
\begin{subequations}
\be
\Hb=\blambda+\eta\hsp(\Ib-\blambda),
\ee
\be
\Vb=N\Hb^{-2} \blambda,\also
\Bb=\sqrt{S}\,\Hb^{-1}(\Ib-\blambda),
\ee
\end{subequations}
we handily rewrite the ``full'' version of eq.~(\ref{msefinal}) in two
spatial variables as the error covariance matrix
\be
\langle\epsilon(\rhat)\epsilon(\rhat')\rangle=
\begg\T\!\!\left(\Vb+\eta^2\Bb^2\right)\!\begg
.
\ee
We subdivide the matrix with Slepian functions into the
truncated set of the best-concentrated $\alpha=1\rar J$ and the
complementary set of remaining $\alpha=J+1\rar\Lpot$ functions,
as follows   
\be
\begg=
\big(\;\beggubar\T \;\; \beggbar\T\big)\T,
\ee
and similarly separate the eigenvalues, writing
\begin{subequations}
\ber
\blambdabar&=&\diag\left(\;\lambda_1\;\cdots\;\lambda_{J}\;\right),\\
\blambdaubar&=&\diag\left(\;\lambda_{J+1}\;\cdots\;\lambda_{\Lpot}\right)
.
\eer
\end{subequations}
Likewise, the identity matrix is split into two parts, $\Ibbar$ and
$\Ibubar$. If we now also redefine 
\begin{subequations}
\be
\Wbbar=N\blambdabar^{-1},\also
\Cbbar=\sqrt{S}\,\Ibbar,
\ee
\be
\Wbubar=N\blambdaubar^{-1},\also
\Cbubar=\sqrt{S}\,\Ibubar,
\ee
\end{subequations}
the equivalent version of eq.~(\ref{SGmsefinal}) is readily
transformed into the full spatial error covariance matrix
\be
\langle\epsilon(\rhat)\epsilon(\rhat')\rangle=
\beggubar\T\Wbubar\hsp\beggubar
+
\beggbar\T\Cbbar^2\hsp\beggbar
.
\ee
In selecting the Slepian basis we have thus successfully separated the
effect of the variance and the bias on the mean-squared reconstruction
error of a noisily observed signal. If the region of observation is a
contiguous closed domain~$R\subset\Omega$ and the truncation should
take place at the Shannon number~$J=N\DDD$, we have thereby identified
the variance as due to \textit{noise} in the region where data are
available, and the bias to \textit{signal} neglected in the truncated
expansion --- which, in the proper Slepian basis, corresponds to the
regions over which no observations exist. In practice, the truncation
will happen at some~$J$ that depends on the signal-to-noise
ratio~\cite[]{Simons+2006b} and/or on computational
considerations~\cite[]{Slobbe+2012}.

Finally, we shall also apply the notions of discretely acquired data
to the solutions of problem~(ii), below.

\ssec{Problem~(ii), Revisited}

We need two more pieces of notation in order to rewrite the
expressions for the spectral estimates~(\ref{SestSP})
and~(\ref{SMTdef}) in the ``pixel-basis''. First we construct the
$M\times M$-dimensional symmetric spatial matrix collecting the
fixed-degree Legendre polynomials evaluated at the angular distances
between all pairs of observations points, 
\be
\bP_l=\tlofp
\left(\barray{ccccc}
P_l(\rhat_1\cdot\rhat_1) & \cdots     & P_l(\rhat_1\cdot\rhat_j)  & \cdots &  P_l(\rhat_1\cdot\rhat_M)\\
                &            & \vdots           &        &       \\
\cdots          &            & P_l(\rhat_i\cdot\rhat_j)    &        &\cdots \\
                &            & \vdots           &        &       \\
P_l(\rhat_M\cdot\rhat_1) & \cdots     & P_l(\rhat_M\cdot\rhat_j)  & \cdots &P_l(\rhat_M\cdot\rhat_M)       
\earray
\right)
.
\ee
The elements of~$\bP_l$ are thus
$\sum_{m=-l}^l\Ylm(\rhat_i)\Ylm(\rhat_j)$,  by the addition
theorem, eq.~(\ref{additionSH}). And finally, we
define~$\bG_l^{\alpha}$, the $M\times M$ symmetric matrix with
elements given by  
\be \label{Kpix}
\left(\bG_l^{\alpha}\right)_{ij}=
\left(\frac{2l+1}{4\pi}\right)g_{\alpha}(\br_i)P_l(\br_i\cdot\br_j)
g_{\alpha}(\br_{j}).
\ee

\sssec{The Spherical Periodogram}

The expression equivalent to eq.~(\ref{SestSP}) is now written as
\be \label{SestSP2}
\hat{S}_l\SP=\left(\frac{4\pi}{A}\right)
\frac{(\Delta\Omega)^2}{2l+1}\,
\bd\T\bP_l\,\bd,
\ee % \left[-\tr(\bN\bP_l)\right]
whereby the column vector~$\bd$ contains the sampled data as in the
notation for eq.~(\ref{beff}). This lends itself easily to
computation, and the statistics of
eqs~(\ref{SexpecSP})--(\ref{moderSP}) hold, approximately, for
sufficiently densely sampled data.  

\sssec{The Spherical Multitaper Estimate}

Finally, the expression equivalent to eq.~(\ref{SMTdef}) becomes
\be \label{Salphadef2}
\hat{S}_l\MT=\sum_{\alpha=1}^{\Lpot}\lambda_\alpha
\left(\frac{4\pi}{N\DDD}\right)\frac{(\Delta\Omega)^2}{2l+1}\,
\bd\T\bG_l^{\alpha}\bd.
\ee
Both eqs~(\ref{SestSP2}) and~(\ref{Salphadef2}) are quadratic forms,
earning them the nickname ``quadratic spectral
estimators''~\cite[]{Mullis+91}. The key difference with the
maximum-likelihood estimator popular in
cosmology~\cite[]{Bond+98,Oh+99,Hinshaw+2003}, which can also be written
as a quadratic form~\cite[]{Tegmark97b}, is that neither~$\bP_l$
nor~$\bG_l^\alpha$ depend on the unknown spectrum itself, and can be
easily precomputed. In contrast, maximum-likelihood estimation is
inherently non-linear, requiring iteration to converge to the most
probable estimate of the power spectral density~\cite[]{Dahlen+2008}. As
such, given a pixel grid, a region of interest~$R$ and a
bandwidth~$L$, eq.~(\ref{Salphadef2}) produces a consistent localized
multitaper power-spectral estimate in one step.

The estimate~(\ref{Salphadef2}) has the statistical properties that we
listed earlier as eqs~(\ref{SMTexpec})--(\ref{Gammafin}). These
continue to hold when the data pixelization is fine enough to have
integral expressions of the kind~(\ref{pixelsum}) be exact.  As
mentioned before, for completely irregularly and potentially
non-densely distributed discrete data on the sphere, ``generalized''
Slepian functions~\cite[]{Bronez88} could be constructed specifically
for the purpose of their power spectral estimation, and used to build
the operator~(\ref{Kpix}). 

\section{Conclusions}

What is the information contained in a bandlimited set of scientific
observations made over an incomplete, e.g. temporally or spatially
limited sampling domain? How can this ``information'', e.g. an
estimate of the signal itself, or of its energy density, be
determined from noisy data, and how shall it be represented? These
seemingly age-old fundamental questions, which have implications
beyond the scientific~\cite[]{Slepian76}, had been solved --- some say,
by conveniently ignoring them --- heuristically, by engineers, well
before receiving their first satisfactory answers given in the
theoretical treatment by Slepian, Landau and
Pollak~\cite[]{Slepian+61,Landau+61,Landau+62}; first for ``continuous''
time series, later generalized to the multidimensional and discrete
cases~\cite[]{Slepian64,Slepian78,Bronez88}.  By the ``Slepian
functions'' in the title of this contribution, we have lumped
together all functions that are ``spatiospectrally'' concentrated,
quadratically, in the original sense of Slepian. In one dimension,
these are the ``prolate spheroidal functions'' whose popularity is as
enduring as their utility. In two Cartesian dimensions, and on the
surface of the unit sphere, both scalar and vectorial, their time for
applications in geomathematics has come.

The answers to the questions posed above are as ever relevant for the
geosciences of today. There, we often face the additional
complications of irregularly shaped study domains, scattered
observations of noise-contaminated potential fields, perhaps collected
from an altitude above the source by airplanes or satellites, and an
acquisition and model-space geometry that is rarely if ever
non-symmetric. Thus the Slepian functions are especially suited for
geoscientific applications and to study any type of geographical
information, in general.

Two problems that are of particular interest in the geosciences, but
also further afield, are how to form a statistically ``optimal''
estimate of the signal giving rise to the data, and how to estimate
the power spectral density of such signal. The first, an inverse
problem that is linear in the data, applies to forming mass flux
estimates from time-variable gravity, e.g. by the \textsc{grace}
mission~\cite[]{Harig+2012}, or to the characterization of the
terrestrial or planetary magnetic fields by satellites such as
\textsc{champ}, \textsc{swarm} or \textsc{mgs}. The second, which is
quadratic in the data, is of interest in studying the statistics of
the Earth's or planetary topography and magnetic
fields~\cite[]{Lewis+2012,Beggan+2013}, and especially for the
cross-spectral analysis of gravity and topography~\cite[]{Wieczorek2008},
which can yield important clues about the internal structure of the
planets. The second problem is also of great interest in cosmology,
where missions such as \textsc{wmap} and \textsc{planck} are mapping
the cosmic microwave background radiation, which is best modeled
spectrally to constrain models of the evolution of our universe.

Slepian functions, as we have shown by focusing on the scalar case in
spherical geometry, provide the mathematical framework to solve such
problems. They are a convenient and easily obtained doubly-orthogonal
mathematical basis in which to express, and thus by which to recover,
signals that are geographically localized, or incompletely (and
noisily) observed. For this they are much better suited than the
traditional Fourier or spherical harmonic bases, and they are more
``geologically intuitive'' than wavelet bases in retaining a firm
geographic footprint and preserving the traditional notions of
frequency or spherical harmonic degree. They are furthermore extremely
performant as data tapers to regularize the inverse problem of power
spectral density determination from noisy and patchy observations,
which can then be solved satisfactorily without costly
iteration. Finally, by the interpretation of the Slepian
functions as their limiting cases, much can be learned about the
statistical nature of such inverse problems when the data provided are
themselves scattered within a specific areal region of study. 

%%%%%%%%%%%%%%%%%%%%%%%%%%%%%%%%%%%%%%%%%%%%%%%%%%%%%%%%%%%%%%%
\acknowledgments

We are indebted to Tony Dahlen (1942--2007), Mark Wieczorek and Volker
Michel for many enlightening discussions over the years. Dong V.~Wang
aided with the calculations of the Cartesian case, and Liying Wei
contributed to the development of the vectorial case. Yoel Shkolnisky
pointed us to the symmetric relations~(\ref{SE})--(\ref{Telements}),
and Korn\'el Jahn shared a preprint of his most recent paper.  Financial 
support for this work was provided by the U.~S.~National Science
Foundation under Grants EAR-0105387, EAR-0710860, EAR-1014606, and
EAR-1150145, by the Universit\'e Paris Diderot--Paris~7 and the
Institut de Physique du Globe de Paris in St.~Maur-des-Foss\'es, the
Ulrich Schmucker Memorial Trust and the Swiss National Science
Foundation. Computer algorithms are made available on
\url{www.frederik.net}.

%%%%%%%%%%%%%%%%%%%%%%%%%%%%%%%%%%%%%%%%%%%%%%%%%%%%%%%%%%%%%
%%%%% References %%%%%
\bibliography{/u/fjsimons/BIBLIO/bib}

\begin{thebibliography}{101}
\expandafter\ifx\csname natexlab\endcsname\relax\def\natexlab#1{#1}\fi

\bibitem[Albertella et~al.(1999)Albertella, Sans{\`o}, \&
  Sneeuw]{Albertella+99}
Albertella, A., Sans{\`o}, F. \& Sneeuw, N., 1999.
Band-limited functions on a bounded spherical domain: the {S}lepian problem on
  the sphere, {\it J.~Geodesy\/}, {\bf 73}, 436--447.

\bibitem[Amirbekyan et~al.(2008)Amirbekyan, Michel, \&
  Simons]{Amirbekyan+2008b}
Amirbekyan, A., Michel, V. \& Simons, F.~J., 2008.
Parameterizing surface-wave tomopgraphic models with harmonic spherical
  splines, {\it Geophys.~J.~Int.\/}, {\bf 174}(2), 617--628, doi:
  10.1111/j.1365--246X.2008.03809.x.

\bibitem[Aster et~al.(2005)Aster, Borchers, \& Thurber]{Aster+2005a}
Aster, R.~C., Borchers, B. \& Thurber, C.~H., 2005.
{\it Parameter Estimation {a}nd Inverse Problems\/}, vol.~90 of {\bf
  International Geophysics Series}, Elsevier Academic Press, San Diego, Calif.

\bibitem[Beggan et~al.(2013)Beggan, Saarim\"aki, Whaler, \&
  Simons]{Beggan+2013}
Beggan, C.~D., Saarim\"aki, J., Whaler, K.~A. \& Simons, F.~J., 2013.
Spectral and spatial decomposition of lithospheric magnetic field models using
  spherical {S}lepian functions, {\it Geophys.~J.~Int.\/}, {\bf 193}(1),
  136--148, 10.1093/gji/ggs122.

\bibitem[Bendat \& Piersol(2000)]{Bendat+2000}
Bendat, J.~S. \& Piersol, A.~G., 2000.
{\it Random data: {A}nalysis {a}nd Measurement Procedures\/}, John Wiley, New
  York, 3rd edn.

\bibitem[Blanco et~al.(1997)Blanco, Fl{\'o}rez, \& Bermejo]{Blanco+97}
Blanco, M.~A., Fl{\'o}rez, M. \& Bermejo, M., 1997.
Evaluation of the rotation matrices in the basis of real spherical harmonics,
  {\it J.~Mol.~Struct. (Theochem)\/}, {\bf 419}, 19--27.

\bibitem[Bond et~al.(1998)Bond, Jaffe, \& Knox]{Bond+98}
Bond, J.~R., Jaffe, A.~H. \& Knox, L., 1998.
Estimating the power spectrum of the cosmic microwave background, {\it
  Phys.~Rev.\ D\/}, {\bf 57}(4), 2117--2137.

\bibitem[Bronez(1988)]{Bronez88}
Bronez, T.~P., 1988.
Spectral estimation of irregularly sampled multidimensional processes by
  generalized prolate spheroidal sequences, {\it IEEE
  Trans.~Acoust.~Speech~Signal~Process.\/}, {\bf 36}(12), 1862--1873.

\bibitem[Chave et~al.(1987)Chave, Thomson, \& Ander]{Chave+87}
Chave, A.~D., Thomson, D.~J. \& Ander, M.~E., 1987.
On the robust estimation of power spectra, coherences, and transfer functions,
  {\it J.~Geophys.~Res.\/}, {\bf 92}(B1), 633--648.

\bibitem[Cohen(1989)]{Cohen89}
Cohen, L., 1989.
Time-frequency distributions --- {A} review, {\it Proc.~IEEE\/}, {\bf 77}(7),
  941--981.

\bibitem[Cox \& Hinkley(1974)]{Cox+74}
Cox, D.~R. \& Hinkley, D.~V., 1974.
{\it Theoretical Statistics\/}, Chapman and Hall, London, UK.

\bibitem[Dahlen \& Simons(2008)]{Dahlen+2008}
Dahlen, F.~A. \& Simons, F.~J., 2008.
Spectral estimation on a sphere in geophysics and cosmology, {\it
  Geophys.~J.~Int.\/}, {\bf 174}, 774--807, doi:
  10.1111/j.1365--246X.2008.03854.x.

\bibitem[Dahlen \& Tromp(1998)]{Dahlen+98}
Dahlen, F.~A. \& Tromp, J., 1998.
{\it Theoretical Global Seismology\/}, Princeton Univ.~Press, Princeton, N.~J.

\bibitem[Daubechies(1988)]{Daubechies88a}
Daubechies, I., 1988.
Time-frequency localization operators: {A} geometric phase space approach, {\it
  IEEE Trans.~Inform.~Theory\/}, {\bf 34}, 605--612.

\bibitem[Daubechies(1990)]{Daubechies90}
Daubechies, I., 1990.
The wavelet transform, time-frequency localization and signal analysis, {\it
  IEEE Trans.~Inform.~Theory\/}, {\bf 36}(5), 961--1005.

\bibitem[Daubechies(1992)]{Daubechies92}
Daubechies, I., 1992.
{\it Ten Lectures on Wavelets\/}, vol.~61 of {\bf CBMS-NSF Regional Conference
  Series in Applied Mathematics}, Society for Industrial \& Applied
  Mathematics, Philadelphia, Penn.

\bibitem[Daubechies \& Paul(1988)]{Daubechies+88}
Daubechies, I. \& Paul, T., 1988.
Time-frequency localisation operators --- {A} geometric phase space approach:
  {II}. {T}he use of dilations, {\it Inv.~Probl.\/}, {\bf 4}(3), 661--680.

\bibitem[de~Villiers et~al.(2003)de~Villiers, Marchaud, \&
  Pike]{DeVilliers+2003}
de~Villiers, G.~D., Marchaud, F. B.~T. \& Pike, E.~R., 2003.
Generalized {G}aussian quadrature applied to an inverse problem in antenna
  theory: {II.} {T}he two-dimensional case with circular symmetry, {\it Inverse
  Problems\/}, {\bf 19}, 755--778.

\bibitem[Donoho \& Stark(1989)]{Donoho+89}
Donoho, D.~L. \& Stark, P.~B., 1989.
Uncertainty principles and signal recovery, {\it SIAM J.~Appl.~Math.\/}, {\bf
  49}(3), 906--931.

\bibitem[Edmonds(1996)]{Edmonds96}
Edmonds, A.~R., 1996.
{\it Angular Momentum in Quantum Mechanics\/}, Princeton Univ.~Press,
  Princeton, N.J.

\bibitem[Eshagh(2009)]{Eshagh2009a}
Eshagh, M., 2009.
Spatially restricted integrals in gradiometric boundary value problems, {\it
  Artif.~Sat.\/}, {\bf 44}(4), 131--148, doi: 10.2478/v10018--009--0025--4.

\bibitem[Flandrin(1998)]{Flandrin98}
Flandrin, P., 1998.
{\it Temps-{F}r{\'e}quence\/}, Herm{\`e}s, Paris, 2nd edn.

\bibitem[Freeden(2010)]{Freeden2010}
Freeden, W., 2010, Geomathematics: Its role, its aim, and its potential, in
  {\em Handbook of Geomathematics\/}, edited by W.~Freeden, M.~Z. Nashed, \&
  T.~Sonar, chap.~1, pp. 3--42, doi: 10.1007/978--3--642--01546--5\_1,
  Springer, Heidelberg, Germany.

\bibitem[Freeden \& Schreiner(2009)]{Freeden+2009}
Freeden, W. \& Schreiner, M., 2009.
{\it Spherical Functions of Mathematical Geosciences: {A} Scalar, Vectorial,
  and Tensorial Setup\/}, Springer, Berlin.

\bibitem[Freeden \& Schreiner(2010)]{Freeden+2010}
Freeden, W. \& Schreiner, M., 2010, Special functions in mathematical
  geosciences: {A}n attempt at a categorization, in {\em Handbook of
  Geomathematics\/}, edited by W.~Freeden, M.~Z. Nashed, \& T.~Sonar, chap.~31,
  pp. 925--948, doi: 10.1007/978--3--642--01546--5\_31, Springer, Heidelberg,
  Germany.

\bibitem[Freeden \& Windheuser(1997)]{Freeden+97}
Freeden, W. \& Windheuser, U., 1997.
Combined spherical harmonic and wavelet expansion --- {A} future concept in
  {E}arth's gravitational determination, {\it Appl.~Comput.~Harmon.~Anal.\/},
  {\bf 4}, 1--37.

\bibitem[Freeden et~al.(1998)Freeden, Gervens, \& Schreiner]{Freeden+98c}
Freeden, W., Gervens, T. \& Schreiner, M., 1998.
{\it Constructive Approximation {o}n {t}he Sphere\/}, Clarendon Press, Oxford,
  UK.

\bibitem[Gerhards(2011)]{Gerhards2011}
Gerhards, C., 2011.
Spherical decompositions in a global and local framework: theory and an
  application to geomagnetic modeling, {\it Intern.~J.~Geomath.\/}, {\bf 1}(2),
  205--256, doi: 10.1007/s13137--010--0011--9.

\bibitem[Gilbert \& Slepian(1977)]{Gilbert+77}
Gilbert, E.~N. \& Slepian, D., 1977.
Doubly orthogonal concentrated polynomials, {\it SIAM J.~Math.~Anal.\/}, {\bf
  8}(2), 290--319.

\bibitem[Grafarend et~al.(2010)Grafarend, Klapp, \& Martinec]{Grafarend+2010}
Grafarend, E.~W., Klapp, M. \& Martinec, Z., 2010, Spacetime modeling of the
  {E}arth's gravity field by ellipsoidal harmonics, in {\em Handbook of
  Geomathematics\/}, edited by W.~Freeden, M.~Z. Nashed, \& T.~Sonar, chap.~7,
  pp. 159--252, doi: 10.1007/978--3--642--01546--5\_7, Springer, Heidelberg,
  Germany.

\bibitem[Grishchuk \& Martin(1997)]{Grishchuk+97}
Grishchuk, L.~P. \& Martin, J., 1997.
Best unbiased estimates for the microwave background anisotropies, {\it
  Phys.~Rev.\ D\/}, {\bf 56}(4), 1924--1938.

\bibitem[Gr{\"u}nbaum(1981)]{Grunbaum81a}
Gr{\"u}nbaum, F.~A., 1981.
Eigenvectors of a {T}oeplitz matrix: discrete version of the prolate spheroidal
  wave functions, {\it SIAM J.~Alg.~Disc.~Meth.\/}, {\bf 2}(2), 136--141.

\bibitem[Han \& Simons(2008)]{Han+2008a}
Han, S.-C. \& Simons, F.~J., 2008.
Spatiospectral localization of global geopotential fields from the {G}ravity
  {R}ecovery and {C}limate {E}xperiment {(GRACE)} reveals the coseismic gravity
  change owing to the 2004 {S}umatra-{A}ndaman earthquake, {\it
  J.~Geophys.~Res.\/}, {\bf 113}, B01405, doi: 10.1029/2007JB004927.

\bibitem[Hanssen(1997)]{Hanssen97}
Hanssen, A., 1997.
Multidimensional multitaper spectral estimation, {\it Signal Process.\/}, {\bf
  58}, 327--332.

\bibitem[Harig \& Simons(2012)]{Harig+2012}
Harig, C. \& Simons, F.~J., 2012.
Mapping {G}reenland's mass loss in space and time, {\it
  Proc.~Natl.~Acad.~Sc.\/}, {\bf 109}(49), 19934--19937, doi:
  10.1073/pnas.1206785109.

\bibitem[Hauser \& Peebles(1973)]{Hauser+73}
Hauser, M.~G. \& Peebles, P. J.~E., 1973.
Statistical analysis of catalogs of extragalactic objects. {II}. {T}he {A}bell
  catalog of rich clusters, {\it Astroph.~J.\/}, {\bf 185}, 757--785.

\bibitem[Hesse et~al.(2010)Hesse, Sloan, \& Womersley]{Hesse+2010}
Hesse, K., Sloan, I.~H. \& Womersley, R.~S., 2010, Numerical integration on the
  sphere, in {\em Handbook of Geomathematics\/}, edited by W.~Freeden, M.~Z.
  Nashed, \& T.~Sonar, chap.~40, pp. 1187--1219, doi:
  10.1007/978--3--642--01546--5\_40, Springer, Heidelberg, Germany.

\bibitem[Hinshaw et~al.(2003)Hinshaw, Spergel, Verde, Hill, Meyer, Barnes,
  Bennett, Halpern, Jarosik, Kogut, Komatsu, Limon, Page, Tucker, Weiland,
  Wollack, \& Wright]{Hinshaw+2003}
Hinshaw, G., Spergel, D.~N., Verde, L., Hill, R.~S., Meyer, S.~S., Barnes, C.,
  Bennett, C.~L., Halpern, M., Jarosik, N., Kogut, A., Komatsu, E., Limon, M.,
  Page, L., Tucker, G.~S., Weiland, J.~L., Wollack, E. \& Wright, E.~L., 2003.
First-year {\textit{{w}ilkinson {m}icrowave {a}nisotropy {p}robe}}
  {\textit{(wmap)}} observations: {T}he angular power spectrum, {\it
  Astroph.~J.~Supp.~Ser.\/}, {\bf 148}, 135--159.

\bibitem[Hivon et~al.(2002)Hivon, G{\'o}rski, Netterfield, Crill, Prunet, \&
  Hansen]{Hivon+2002}
Hivon, E., G{\'o}rski, K.~M., Netterfield, C.~B., Crill, B.~P., Prunet, S. \&
  Hansen, F., 2002.
{MASTER} of the cosmic microwave background anisotropy power spectrum: {A} fast
  method for statistical analysis of large and complex cosmic microwave
  background data sets, {\it Astroph.~J.\/}, {\bf 567}, 2--17.

\bibitem[Ilk(1983)]{Ilk83}
Ilk, K.~H., 1983.
Ein {B}eitrag zur {D}ynamik ausgedehnter {K}\"orper:
  {G}ravitationswechselwirkung, {\it Deutsche Geod\"atische Kommission\/}, {\bf
  C}(288).

\bibitem[Jahn \& Bokor(2012)]{Jahn+2012}
Jahn, K. \& Bokor, N., 2012.
Vector slepian basis functions with optimal energy concentration in high
  numerical aperture focusing, {\it Optics Comm.\/}, {\bf 285}, 2028--2038,
  doi: 10.1016/j.optcom.2011.11.107.

\bibitem[Jahn \& Bokor(2013{\natexlab{a}})]{Jahn+2013a}
Jahn, K. \& Bokor, N., 2013.
Solving the inverse problem of high numerical aperture focusing using vector
  slepian harmonics and vector {S}lepian multipole fields, {\it Optics
  Comm.\/}, {\bf 288}, 13--16, doi: 10.1016/j.optcom.2012.09.051.

\bibitem[Jahn \& Bokor(2013{\natexlab{b}})]{Jahn+2013b}
Jahn, K. \& Bokor, N., 2013.
Revisiting the concentration problem of vector fields within a spherical cap: a
  commuting differential operator solution, {\it
  Appl.~Comput.~Harmon.~Anal.\/}, pp. submitted, preprint available as
  http://arxiv.org/abs/1302.5261.

\bibitem[Jones(1963)]{Jones63}
Jones, R.~H., 1963.
Stochastic processes on a sphere, {\it Ann.~Math.~Stat.\/}, {\bf 34}(1),
  213--218.

\bibitem[Kaula(1967)]{Kaula67a}
Kaula, W.~M., 1967.
Theory of statistical analysis of data distributed over a sphere, {\it
  Rev.~Geophys.\/}, {\bf 5}(1), 83--107.

\bibitem[Kennedy \& Sadeghi(2013)]{Kennedy+2013}
Kennedy, R.~A. \& Sadeghi, P., 2013.
{\it Hilbert Space Methods in Signal Processing\/}, Cambridge Univ.~Press,
  Cambridge, UK.

\bibitem[Knox(1995)]{Knox95}
Knox, L., 1995.
Determination of inflationary observables by cosmic microwave background
  anisotropy experiments, {\it Phys.~Rev.\ D\/}, {\bf 52}(8), 4307--4318.

\bibitem[Landau(1965)]{Landau65}
Landau, H.~J., 1965.
On the eigenvalue behavior of certain convolution equations, {\it
  Trans.~Am.~Math.~Soc.\/}, {\bf 115}, 242--256.

\bibitem[Landau \& Pollak(1961)]{Landau+61}
Landau, H.~J. \& Pollak, H.~O., 1961.
Prolate spheroidal wave functions, {F}ourier analysis and uncertainty --- {II},
  {\it Bell Syst.~Tech.~J.\/}, {\bf 40}(1), 65--84.

\bibitem[Landau \& Pollak(1962)]{Landau+62}
Landau, H.~J. \& Pollak, H.~O., 1962.
Prolate spheroidal wave functions, {F}ourier analysis and uncertainty ---
  {III}: {T}he dimension of the space of essentially time- and band-limited
  signals, {\it Bell Syst.~Tech.~J.\/}, {\bf 41}(4), 1295--1336.

\bibitem[Lewis \& Simons(2012)]{Lewis+2012}
Lewis, K.~W. \& Simons, F.~J., 2012.
Local spectral variability and the origin of the {M}artian crustal magnetic
  field, {\it Geophys.~Res.~Lett.\/}, {\bf 39}, L18201, doi:
  10.1029/2012GL052708.

\bibitem[Mallat(1998)]{Mallat98}
Mallat, S., 1998.
{\it A Wavelet Tour {o}f Signal Processing\/}, Academic Press, San Diego,
  Calif.

\bibitem[Maniar \& Mitra(2005)]{Maniar+2005}
Maniar, H. \& Mitra, P.~P., 2005.
The concentration problem for vector fields, {\it Int.~J.~Bioelectromagn.\/},
  {\bf 7}(1), 142--145.

\bibitem[Martinec(2010)]{Martinec2010}
Martinec, Z., 2010, The forward and adjoint methods of global electromagnetic
  induction for {CHAMP} magnetic data, in {\em Handbook of Geomathematics\/},
  edited by W.~Freeden, M.~Z. Nashed, \& T.~Sonar, chap.~19, pp. 565--624, doi:
  10.1007/978--3--642--01546--5\_19, Springer, Heidelberg, Germany.

\bibitem[Menke(1989)]{Menke89}
Menke, W., 1989.
{\it Geophysical Data Analysis: {D}iscrete Inverse Theory\/}, vol.~45 of {\bf
  International Geophysics Series}, Academic Press, San Diego, Calif., {R}ev.
  edn.

\bibitem[Messiah(2000)]{Messiah2000}
Messiah, A., 2000.
{\it Quantum Mechanics\/}, Dover, New York.

\bibitem[Michel(2010)]{Michel2010}
Michel, V., 2010, Tomography: {P}roblems and multiscale solutions, in {\em
  Handbook of Geomathematics\/}, edited by W.~Freeden, M.~Z. Nashed, \&
  T.~Sonar, chap.~32, pp. 949--972, doi: 10.1007/978--3--642--01546--5\_32,
  Springer, Heidelberg, Germany.

\bibitem[Mitra \& Maniar(2006)]{Mitra+2006}
Mitra, P.~P. \& Maniar, H., 2006.
Concentration maximization and local basis expansions ({LBEX}) for linear
  inverse problems, {\it IEEE Trans.~Biomed Eng.\/}, {\bf 53}(9), 1775--1782.

\bibitem[Mullis \& Scharf(1991)]{Mullis+91}
Mullis, C.~T. \& Scharf, L.~L., 1991, Quadratic estimators {o}f {t}he power
  spectrum, in {\em Advances {i}n Spectrum Analysis {a}nd Array Processing\/},
  edited by S.~Haykin, vol.~1, chap.~1, pp. 1--57, Prentice-Hall, Englewood
  Cliffs, N.~J.

\bibitem[Nashed \& Walter(1991)]{Nashed+91}
Nashed, M.~Z. \& Walter, G.~G., 1991.
General sampling theorems for functions in {R}eproducing {K}ernel {H}ilbert
  {S}paces, {\it Math.~Control Signals Syst.\/}, {\bf 4}, 363--390.

\bibitem[Oh et~al.(1999)Oh, Spergel, \& Hinshaw]{Oh+99}
Oh, S.~P., Spergel, D.~N. \& Hinshaw, G., 1999.
An efficient technique to determine the power spectrum from cosmic microwave
  background sky maps, {\it Astroph.~J.\/}, {\bf 510}, 551--563.

\bibitem[Olsen et~al.(2010)Olsen, Hulot, \& Sabaka]{Olsen+2010}
Olsen, N., Hulot, G. \& Sabaka, T.~J., 2010, Sources of the geomagnetic field
  and the modern data that enable their investigation, in {\em Handbook of
  Geomathematics\/}, edited by W.~Freeden, M.~Z. Nashed, \& T.~Sonar, chap.~5,
  pp. 105--124, doi: 10.1007/978--3--642--01546--5\_5, Springer, Heidelberg,
  Germany.

\bibitem[Paul(1978)]{Paul78}
Paul, M.~K., 1978.
Recurrence relations for integrals of associated {L}egendre functions, {\it
  Bull.~G{\'e}od.\/}, {\bf 52}, 177--190.

\bibitem[Peebles(1973)]{Peebles73}
Peebles, P. J.~E., 1973.
Statistical analysis of catalogs of extragalactic objects. {I}. {T}heory, {\it
  Astroph.~J.\/}, {\bf 185}, 413--440.

\bibitem[Percival \& Walden(1993)]{Percival+93}
Percival, D.~B. \& Walden, A.~T., 1993.
{\it Spectral analysis {f}or physical applications, multitaper {a}nd
  conventional univariate techniques\/}, Cambridge Univ.~Press, New York.

\bibitem[Plattner \& Simons(2013)]{Plattner+2013}
Plattner, A. \& Simons, F.~J., 2013.
Spatiospectral concentration of vector fields on a sphere, {\it
  Appl.~Comput.~Harmon.~Anal.\/}, p. doi:10.1016/j.acha.2012.12.001.

\bibitem[Plattner et~al.(2012)Plattner, Simons, \& Wei]{Plattner+2012}
Plattner, A., Simons, F.~J. \& Wei, L., 2012, Analysis of real vector fields on
  the sphere using {S}lepian functions, in {\em 2012 IEEE Statistical Signal
  Processing Workshop (SSP'12)\/}, IEEE.

\bibitem[Riedel \& Sidorenko(1995)]{Riedel+95}
Riedel, K.~S. \& Sidorenko, A., 1995.
Minimum bias multiple taper spectral estimation, {\it IEEE
  Trans.~Signal~Process.\/}, {\bf 43}(1), 188--195.

\bibitem[Sabaka et~al.(2010)Sabaka, Hulot, \& Olsen]{Sabaka+2010}
Sabaka, T.~J., Hulot, G. \& Olsen, N., 2010, Mathematical properties relevant
  to geomagnetic field modeling, in {\em Handbook of Geomathematics\/}, edited
  by W.~Freeden, M.~Z. Nashed, \& T.~Sonar, chap.~17, pp. 503--538, doi:
  10.1007/978--3--642--01546--5\_17, Springer, Heidelberg, Germany.

\bibitem[Schuster(1898)]{Schuster1898}
Schuster, A., 1898.
An investigation of hidden periodicities with application to a supposed 26-day
  period of meteorological phenomena, {\it Terr.~Magn.\/}, {\bf 3}, 13--41.

\bibitem[Sheppard \& T\"or\"ok(1997)]{Sheppard+97}
Sheppard, C. J.~R. \& T\"or\"ok, P., 1997.
Efficient calculation of electromagnetic diffraction in optical systems using a
  multipole expansion, {\it J.~Mod.~Optics\/}, {\bf 44}(4), 803--818, doi:
  10.1080/09500349708230696.

\bibitem[Shkolnisky(2007)]{Shkolnisky2007}
Shkolnisky, Y., 2007.
Prolate spheroidal wave functions on a disc --- {I}ntegration and approximation
  of two-dimensional bandlimited functions, {\it
  Appl.~Comput.~Harmon.~Anal.\/}, {\bf 22}, 235--256, doi:
  10.1016/j.acha.2006.07.002.

\bibitem[Simons(2010)]{Simons2010}
Simons, F.~J., 2010, Slepian functions and their use in signal estimation and
  spectral analysis, in {\em Handbook of Geomathematics\/}, edited by
  W.~Freeden, M.~Z. Nashed, \& T.~Sonar, chap.~30, pp. 891--923, doi:
  10.1007/978--3--642--01546--5\_30, Springer, Heidelberg, Germany.

\bibitem[Simons \& Dahlen(2006)]{Simons+2006b}
Simons, F.~J. \& Dahlen, F.~A., 2006.
Spherical {S}lepian functions and the polar gap in geodesy, {\it
  Geophys.~J.~Int.\/}, {\bf 166}, 1039--1061, doi:
  10.1111/j.1365--246X.2006.03065.x.

\bibitem[Simons \& Dahlen(2007)]{Simons+2007}
Simons, F.~J. \& Dahlen, F.~A., 2007, A spatiospectral localization approach to
  estimating potential fields on the surface of a sphere from noisy, incomplete
  data taken at satellite altitudes, in {\em Wavelets {XII}\/}, edited by
  D.~{Van de Ville}, V.~K. Goyal, \& M.~Papadakis, vol. 6701, pp. 670117, doi:
  10.1117/12.732406, SPIE.

\bibitem[Simons \& Wang(2011)]{Simons+2011a}
Simons, F.~J. \& Wang, D.~V., 2011.
Spatiospectral concentration in the {C}artesian plane, {\it
  Intern.~J.~Geomath.\/}, {\bf 2}(1), 1--36, doi: 10.1007/s13137--011--0016--z.

\bibitem[Simons et~al.(2000)Simons, Zuber, \& Korenaga]{Simons+2000}
Simons, F.~J., Zuber, M.~T. \& Korenaga, J., 2000.
Isostatic response of the {A}ustralian lithosphere: {E}stimation of effective
  elastic thickness and anisotropy using multitaper spectral analysis, {\it
  J.~Geophys.~Res.\/}, {\bf 105}(B8), 19163--19184, doi: 10.1029/2000JB900157.

\bibitem[Simons et~al.(2006)Simons, Dahlen, \& Wieczorek]{Simons+2006a}
Simons, F.~J., Dahlen, F.~A. \& Wieczorek, M.~A., 2006.
Spatiospectral concentration on a sphere, {\it SIAM Rev.\/}, {\bf 48}(3),
  504--536, doi: 10.1137/S0036144504445765.

\bibitem[Simons et~al.(2009)Simons, Hawthorne, \& Beggan]{Simons+2009b}
Simons, F.~J., Hawthorne, J.~C. \& Beggan, C.~D., 2009, Efficient analysis and
  representation of geophysical processes using localized spherical basis
  functions, in {\em Wavelets~{XIII}\/}, edited by V.~K. Goyal, M.~Papadakis,
  \& D.~{Van de Ville}, vol. 7446, pp. 74460G, doi: 10.1117/12.825730, SPIE.

\bibitem[Slepian(1964)]{Slepian64}
Slepian, D., 1964.
Prolate spheroidal wave functions, {F}ourier analysis and uncertainty --- {IV}:
  {E}xtensions to many dimensions; generalized prolate spheroidal functions,
  {\it Bell Syst.~Tech.~J.\/}, {\bf 43}(6), 3009--3057.

\bibitem[Slepian(1976)]{Slepian76}
Slepian, D., 1976.
On bandwidth, {\it Proc.~IEEE\/}, {\bf 64}(3), 292--300.

\bibitem[Slepian(1978)]{Slepian78}
Slepian, D., 1978.
Prolate spheroidal wave functions, {F}ourier analysis and uncertainty --- {V}:
  {T}he discrete case, {\it Bell Syst.~Tech.~J.\/}, {\bf 57}, 1371--1429.

\bibitem[Slepian(1983)]{Slepian83}
Slepian, D., 1983.
Some comments on {F}ourier analysis, uncertainty and modeling, {\it SIAM
  Rev.\/}, {\bf 25}(3), 379--393.

\bibitem[Slepian \& Pollak(1961)]{Slepian+61}
Slepian, D. \& Pollak, H.~O., 1961.
Prolate spheroidal wave functions, {F}ourier analysis and uncertainty --- {I},
  {\it Bell Syst.~Tech.~J.\/}, {\bf 40}(1), 43--63.

\bibitem[Slepian \& Sonnenblick(1965)]{Slepian+65}
Slepian, D. \& Sonnenblick, E., 1965.
Eigenvalues associated with prolate spheroidal wave functions of zero order,
  {\it Bell Syst.~Tech.~J.\/}, {\bf 44}(8), 1745--1759.

\bibitem[Slobbe et~al.(2012)Slobbe, Simons, \& Klees]{Slobbe+2012}
Slobbe, D.~C., Simons, F.~J. \& Klees, R., 2012.
The spherical {S}lepian basis as a means to obtain spectral consistency between
  mean sea level and the geoid, {\it J.~Geodesy\/}, {\bf 86}(8), 609--628, doi:
  10.1007/s00190--012--0543--x.

\bibitem[Sneeuw \& van Gelderen(1997)]{Sneeuw+97}
Sneeuw, N. \& van Gelderen, M., 1997, The polar gap, in {\em Geodetic boundary
  value problems in view of the one centimeter geoid\/}, edited by F.~Sans{\`o}
  \& R.~Rummel, vol.~65 of {\bf Lecture Notes in Earth Sciences}, pp. 559--568,
  Springer, Berlin.

\bibitem[Tegmark(1997)]{Tegmark97b}
Tegmark, M., 1997.
How to measure {CMB} power spectra without losing information, {\it Phys.~Rev.\
  D\/}, {\bf 55}(10), 5895--5907.

\bibitem[Tegmark et~al.(1997)Tegmark, Taylor, \& Heavens]{Tegmark+97}
Tegmark, M., Taylor, A.~N. \& Heavens, A.~F., 1997.
Karhunen-{L}o{\`e}ve eigenvalue problems in cosmology: {H}ow should we tackle
  large data sets?, {\it Astroph.~J.\/}, {\bf 480}(1), 22--35.

\bibitem[Thomson(1982)]{Thomson82}
Thomson, D.~J., 1982.
Spectrum estimation and harmonic analysis, {\it Proc.~IEEE\/}, {\bf 70}(9),
  1055--1096.

\bibitem[Thomson(2007)]{Thomson2007}
Thomson, D.~J., 2007.
Jackknifing multitaper spectrum estimates, {\it IEEE Signal~Process.~Mag.\/},
  {\bf 20}, 20--30, doi: 0.1109/MSP.2007.4286561.

\bibitem[Thomson \& Chave(1991)]{Thomson+91}
Thomson, D.~J. \& Chave, A.~D., 1991, Jackknifed error estimates {f}or spectra,
  coherences, {a}nd transfer functions, in {\em Advances {i}n Spectrum Analysis
  {a}nd Array Processing\/}, edited by S.~Haykin, vol.~1, chap.~2, pp. 58--113,
  Prentice-Hall, Englewood Cliffs, N.~J.

\bibitem[Tricomi(1970)]{Tricomi70}
Tricomi, F.~G., 1970.
{\it Integral Equations\/}, Interscience, New York, 5th edn.

\bibitem[Varshalovich et~al.(1988)Varshalovich, Moskalev, \&
  Kherso{\'n}skii]{Varshalovich+88}
Varshalovich, D.~A., Moskalev, A.~N. \& Kherso{\'n}skii, V.~K., 1988.
{\it Quantum theory of angular momentum\/}, World Scientific, Singapore.

\bibitem[Wieczorek(2008)]{Wieczorek2008}
Wieczorek, M.~A., 2008.
Constraints on the composition of the martian south polar cap from gravity and
  topography, {\it Icarus\/}, {\bf 196}(2), 506--517, doi:
  10.1016/j.icarus.2007.10.026.

\bibitem[Wieczorek \& Simons(2005)]{Wieczorek+2005}
Wieczorek, M.~A. \& Simons, F.~J., 2005.
Localized spectral analysis on the sphere, {\it Geophys.~J.~Int.\/}, {\bf
  162}(3), 655--675, doi: 10.1111/j.1365--246X.2005.02687.x.

\bibitem[Wieczorek \& Simons(2007)]{Wieczorek+2007}
Wieczorek, M.~A. \& Simons, F.~J., 2007.
Minimum-variance spectral analysis on the sphere, {\it J.~Fourier
  Anal.~Appl.\/}, {\bf 13}(6), 665--692, doi: 10.1007/s00041--006--6904--1.

\bibitem[Wingham(1992)]{Wingham92}
Wingham, D.~J., 1992.
The reconstruction of a band-limited function and its {F}ourier transform from
  a finite number of samples at arbitrary locations by {S}ingular {V}alue
  {D}ecomposition, {\it IEEE Trans.~Signal~Process.\/}, {\bf 40}(3), 559--570,
  doi: 10.1109/78.120799.

\bibitem[Xu(1992)]{Xu92a}
Xu, P., 1992.
Determination of surface gravity anomalies using gradiometric observables, {\it
  Geophys.~J.~Int.\/}, {\bf 110}, 321--332.

\bibitem[Xu(1998)]{Xu98}
Xu, P., 1998.
Truncated {SVD} methods for discrete linear ill-posed problems, {\it
  Geophys.~J.~Int.\/}, {\bf 135}(2), 505--514, doi:
  10.1046/j.1365--246X.1998.00652.x.

\bibitem[Yao(1967)]{Yao67}
Yao, K., 1967.
Application of reproducing kernel {H}ilbert spaces --- {B}andlimited signal
  models, {\it Inform.~Control\/}, {\bf 11}(4), 429--444.

\end{thebibliography}
\bibliographystyle{gji}

\end{document}